\documentclass[
aps,prd,twocolumn,preprintnumbers,superscriptaddress,
nofootinbib,floatfix,
amsmath,amssymb,dvipsnames,
longbibliography
]{revtex4-2}

\usepackage[utf8]{inputenc}
\usepackage{siunitx}
\usepackage{amsfonts}
\usepackage{braket}
\usepackage{mathtools}
\usepackage{cancel}
\usepackage{slashed}
\usepackage{pifont}
\usepackage{soul}
\usepackage{comment}

\usepackage{yhmath}

\usepackage{booktabs,array}
\usepackage{hhline}
\usepackage{dcolumn}
    \newcolumntype{d}[1]{D{.}{.}{#1}}
\usepackage{graphicx}
\usepackage[caption=false]{subfig}
\usepackage{multirow}

\usepackage{dcolumn}
\usepackage{mathtools}
\usepackage{upgreek}
\newcolumntype{T}{D{.}{.}{10}}
\newcolumntype{E}{D{.}{.}{11}}
\newcolumntype{F}{D{.}{.}{5}}

\usepackage[normalem]{ulem}

\usepackage[dvipsnames]{xcolor}
\usepackage{hyperref}
\hypersetup{
    colorlinks=true,
    linkcolor=blue,
    citecolor=blue,
    filecolor=blue,
    urlcolor=blue
}

\usepackage{bm}

\renewcommand{\vec}[1]{{\mathbf{#1}}}

\makeatletter 
\renewcommand\onecolumngrid{
\do@columngrid{one}{\@ne}
\def\set@footnotewidth{\onecolumngrid}
\def\footnoterule{\kern-6pt\hrule width 1.5in\kern6pt}
}
\renewcommand\twocolumngrid{
        \def\footnoterule{
        \dimen@\skip\footins\divide\dimen@\thr@@
        \kern-\dimen@\hrule width.5in\kern\dimen@}
        \do@columngrid{mlt}{\tw@}
}
\makeatother

\begin{document}
\title{Lattice evidence that scalar glueballs are small}

\author{Ryan Abbott}
\affiliation{Center for Theoretical Physics---A Leinweber Institute, Massachusetts Institute of Technology, Cambridge, MA 02139, U.S.A.}
\author{Daniel C. Hackett}
\affiliation{Fermi National Accelerator Laboratory, Batavia, IL 60510, U.S.A.}
\author{Dimitra A. Pefkou}
\affiliation{Department of Physics, University of California, Berkeley, CA 94720, U.S.A}
\affiliation{Nuclear Science Division, Lawrence Berkeley National Laboratory, Berkeley, CA 94720, USA}
\author{Fernando~Romero-López}
\affiliation{Albert Einstein Center, Institute for Theoretical Physics, University of Bern, 3012 Bern, Switzerland}
\author{Phiala E. Shanahan}
\affiliation{Center for Theoretical Physics---A Leinweber Institute, Massachusetts Institute of Technology, Cambridge, MA 02139, U.S.A.}

\begin{abstract}
This work reports the first calculation of the gravitational form factors (GFFs) of the scalar glueball, performed via lattice field theory in Yang-Mills theory at a single lattice spacing. The glueball GFFs are compared with those of other hadrons as determined in previous lattice calculations, providing strong indications that glueballs have a different gluonic structure than typical hadronic states. A mass radius of 0.263(31)~fm is predicted, supporting previous suggestions that the scalar glueball is significantly smaller than other hadrons.
These results point towards a potential smoking-gun characteristic to target by experimental glueball searches.
\end{abstract}

\preprint{MIT-CTP/5907, FERMILAB-PUB-25-0621-T, INT-PUB-25-021}

\maketitle

\textit{Introduction}: Since the inception of quantum chromodynamics (QCD)~\cite{Fritzsch:1972jv}, glueballs have been postulated as hadronic states with purely gluonic degrees of freedom. Decades of experiments in hadron spectroscopy have collected experimental glueball candidates with a variety of allowed quantum numbers, e.g., ${J^{PC} = 0^{++},0^{-+},1^{++},1^{+-},1^{--},2^{++}}$, etc.~\cite{Klempt:2007cp,Crede:2008vw,Chen:2022asf,BESIII:2010gmv,BESIII:2019wkp,BESIII:2023wfi}.  Lattice QCD calculations of hadronic resonances provide valuable guidance towards the identification of putative glueball states~\cite{Teper:1998kw,Morningstar:1999rf,Athenodorou:2016ebg,Athenodorou:2020ani,Sakai:2022zdc}---see Refs.~\cite{Vadacchino:2023vnc,Morningstar:2024vjk} for a review. However, identifying observed hadrons as glueballs or glueball-like states based on spectroscopy remains challenging due to the mixing of hadrons with the same quantum numbers~\cite{Bali:2000vr,Athenodorou:2023ntf}.

Beyond the spectrum, information about the internal structure of hadrons may provide an additional pathway for classification of observed states as glueball-like or non-glueball objects. For example, features such as the momentum fraction carried by gluons may be qualitatively different for glueball and non-glueball states, and could serve as 
evidence of a hadron having predominantly gluonic degrees of freedom. The size of glueballs, which can be defined in various ways, can serve as another way to distinguish them from typical hadrons; model studies~\cite{Schafer:1994fd,Forkel:2000fd,Hou:2002jv} have speculated that the scalar glueball might be more compact than the $\Lambda_{\text{QCD}}$ scale that is thought to define the size of more typical hadrons. This was supported by early lattice studies which, however, relied on a model-dependent Bethe-Salpeter approach to define the radius~\cite{PhysRevD.43.2301,PhysRevLett.69.245,Loan:2006gm}, or were performed in $SU(2)$ Yang-Mills theory~\cite{Tickle:1989gw}. The typical scales found, around 0.2~fm, are also significantly smaller than the radius of the $\sigma$ meson, which is the lightest resonance with the same quantum numbers, as predicted from a dispersive analysis~\cite{Albaladejo:2012te}. 

The \textit{gravitational form factors} (GFFs) of hadrons encode the gluon momentum fraction and distribution of quantities like the energy inside hadrons~\cite{Polyakov:2002yz}, offering a robust way to define their size through, e.g., their mass radius, defined from the root mean square of the energy density. The GFFs are defined from the matrix elements of the \textit{energy-momentum tensor} (EMT) $T^{\mu\nu}$ of QCD~\cite{Lorce:2018egm,Polyakov:2002yz,Polyakov:2018zvc,Burkert:2023wzr}.
Lattice QCD provides a tool to determine these matrix elements, and thus the glueball GFFs. While previous pioneering work~\cite{Chen:2005mg,Tickle:1989gw} has investigated glueball matrix elements using lattice field theory, the GFFs of glueballs have not been previously constrained.

This work presents a first step towards quantitative studies of glueball structure through a calculation of the GFFs of the lowest-lying scalar glueball state\footnote{See Ref.~\cite{Abbott:2024bre} for preliminary results of this work.}, denoted as $G[0^{++}]$, in $SU(3)$ Yang-Mills theory. In a pure gauge theory, the EMT is purely gluonic, defined as
${T^{\mu\nu}= 2\;\text{Tr}[-F^{\mu}_{\alpha}F^{\alpha\nu}+\frac{1}{4}g^{\mu\nu}F^{\alpha\beta}F_{\alpha\beta}]}$,
where $F^{\mu\nu}$ is the gluon field strength tensor. 
The EMT matrix element for the scalar glueball can be decomposed in terms of two GFFs,  $A(t)$ and $D(t)$~\cite{Pagels:1966zza,Hudson:2017xug,Cotogno:2019vjb}:
\begin{equation} \label{eq:ME}
\begin{split}
    \left \langle G[0^{++}](p') \right| T^{\mu\nu} \left| G[0^{++}](p) \right \rangle &= \\
    2 P^{\mu}P^{\nu}A(t) 
    +&\frac{\Delta^{\mu} \Delta^{\nu}-g^{\mu\nu} \Delta^2}{2} D(t) \;,
\end{split}
\end{equation}
where the four-momenta of the incoming and outgoing states are denoted $p$ and $p'$,
$P=(p+p')/2$, $\Delta=p'-p$, and $t=\Delta^2$. Note that $A(0)=1$ follows from the momentum sum rule, while the $D$-term, $D(0)$, is an unconstrained and previously unknown quantity.

\textit{Methodology}: Calculations are performed using a single pure-gauge ensemble defined by the $SU(3)$ Wilson gauge action with $\beta=5.95$. This results in $a=0.098$~fm, using the Sommer parameter to set the scale~\cite{Necco:2001xg,Durr:2006ky}. The lattice geometry is $L^3 \times T = 24^3\times 48$. 
Calculations are performed using $\mathcal{O}(10^7)$ 
configurations generated using $\mathcal{O}(10^5)$ independent streams of heatbath with overrelaxation~\cite{Creutz:1980zw, Cabibbo:1982zn, Kennedy:1985nu, Brown:1987rra, Adler:1987ce}, which were saved every 25 heatbath hits.
Two-point correlation functions are constructed on each configuration using two interpolating operators:
\begin{align} \label{eq:interp}
\begin{split}
    \chi_1(x)  = \frac{1}{4} \sum_{\mu\neq\nu}\text{Re Tr} \, U^2_{\mu\nu}(x), \\  \chi_2(x) = \frac{1}{4}\sum_{\mu\neq\nu}\text{Re Tr} \, U^7_{\mu\nu}(x) \;,
\end{split}
\end{align}
where $\mu,\nu\in\{x,y,z\}$. Here, $U^n_{\mu\nu}$ is an $n \times n$ Wilson loop which extends in the $\mu$ and $\nu$ directions constructed from links stout-smeared~\cite{Morningstar:2003gk} by 3 steps in the spatial directions only. 
It is convenient to define vacuum-subtracted operators of definite three-momentum $\vec{p}$ as:
\begin{equation}
    \chi_i(\vec{p}, t) =  \sum_{\vec{x}} e^{-i \vec{p} \cdot \vec{x}} 
    \bigg[ \chi_i(\vec{x},t) - \braket{\chi_i(\vec{x},t)}\bigg] .
    \label{eq:interp-p}
\end{equation}
The positive-parity $0^{++}$ glueball is the lowest-energy state excited by these operators; the summations over $\mu,\nu$ in Eq.~\eqref{eq:interp} project to the $A_1^+$ (rest frame) or $A_1$ (moving frames) irreducible representations (irreps) of the finite-volume symmetry group, while taking the real part projects to positive charge conjugation quantum numbers. These irreps have maximal overlap with the scalar glueball; however, depending on the momentum frame, the spectrum may also include heavier glueballs with other quantum numbers, e.g.~tensor or pseudoscalar glueballs, or multi-glueball or higher-lying ditorelon states~\cite{Michael:1989vh}.

The spectrum of glueball states is constrained via the generalized eigenvalue problem (GEVP) method \cite{Fox:1981xz,Michael:1982gb,Luscher:1990ck,Blossier:2009kd,Fleming:2023zml,Fischer:2020bgv,Sterman:2000pu}. The GEVP is applied to the $2\times 2$ matrix of momentum-projected two-point functions averaged over all timeslices:
\begin{equation}
C^{2\text{pt}}_{ij}(\vec{p},t) = \frac{1}{T} \sum_{t_0} \braket{ \chi_i(\vec{p},t+t_0) \, \chi_j(\vec{p}, t_0)^\dagger }
\label{eq:2pt-matrix}
\end{equation}
for all $\left|\vec{p}\right|^2\leq 6 (2\pi/L)^2$ on 200 bootstrap ensembles after binning the $\mathcal{O}(10^7)$ configurations into groups of $1000$ consecutive configurations.
Averaging over equivalent momenta yields two-point functions for the 7 distinct $|\vec{p}|^2$; analyzing them in a ``fixed pivot''~\cite{Fox:1981xz,Michael:1982gb,Luscher:1990ck} mode with $t_0=1$ and diagonalization time $t_d=3$ gives
7 sets of weights $w_{ij}{(|\vec{p}|^2)}$ used to construct optimized ground-state interpolating operators:
\begin{equation}
    \chi_{0}(\vec{p},t) = \sum_i w_{0i}{(|\vec{p}|^2)} \chi_i(\vec{p},t) .
    \label{eq:interp-gevp}
\end{equation}
These are used to construct two-point correlation functions as:
\begin{equation}
\begin{split}
    C_{0^{++}}^{2\text{pt}}(\vec{p},t) 
    &= \frac{1}{T} \sum_{t_0} \braket{\chi_{0}(\vec{p},t+t_0) ~ \chi_{0}(\vec{p},t_0)^\dagger } \\
    &= \sum_{i,j} w_{0i}(|\vec{p}|^2) \, C^{2\mathrm{pt}}_{ij}(\vec{p},t) \, w^*_{0j}(|\vec{p}|^2).
\end{split}
    \label{eq:gevp-2pt}
\end{equation}
For each $\left| \vec{p} \right|^2$, the generalized eigenvalue problem is then solved to extract the ground state, which is identified as the scalar glueball. Figures showing examples of the two-point correlation functions, and the corresponding analysis, are included in the Supplementary Material (SM).

\begin{figure*} 
\includegraphics[width=0.98\linewidth]{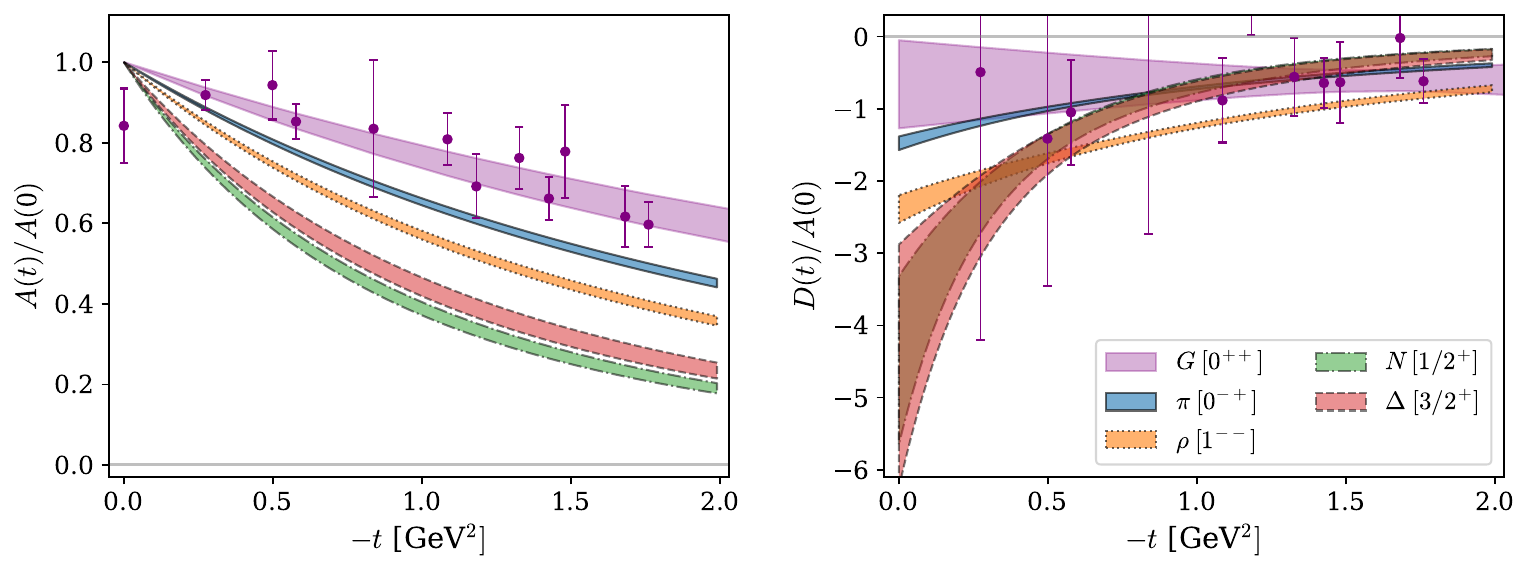}
\caption{
Comparison of the $t$-dependence of the $G[0^{++}]$ glueball GFFs in Yang-Mills theory obtained in this work, and the gluon GFFs of 
four other hadrons---the pion, $\rho$ meson, nucleon, and $\Delta$ baryon, indicated with their $J^P$ quantum numbers---obtained in a $N_f=2+1$ lattice QCD calculation with $m_{\pi}\simeq 450~\text{MeV}$~\cite{Pefkou:2021fni}. 
}
\label{fig:GFFs}
\end{figure*}

\begin{figure*}[t]
\includegraphics[width=0.98\linewidth]{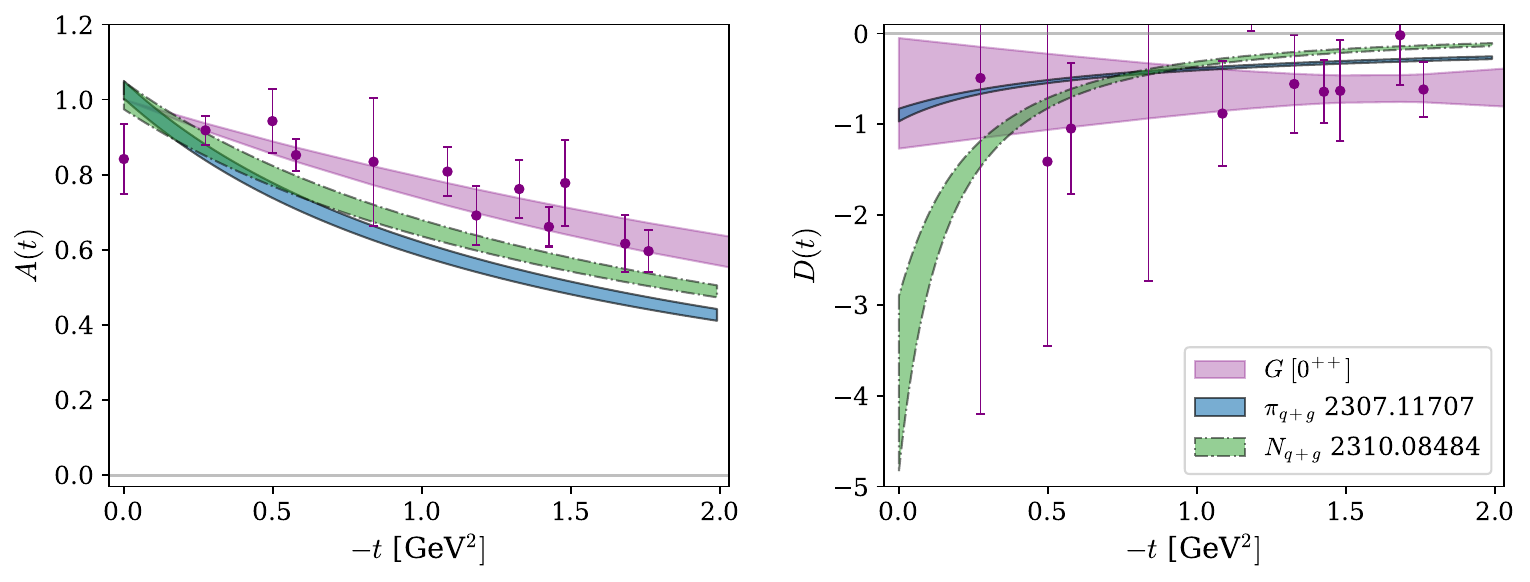}
\caption{
Comparison between the $G[0^{++}]$ glueball GFFs in Yang-Mills theory obtained in this work, the total GFFs of 
the nucleon (using dipole fits) and the pion (using monopole fits) obtained 
with an $N_f=2+1$ QCD ensemble with $m_{\pi}=170~\text{MeV}$~\cite{Hackett:2023rif,Hackett:2023nkr}.  
}
\label{fig:170}
\end{figure*}

\begin{figure} 
\includegraphics[width=0.98\linewidth]{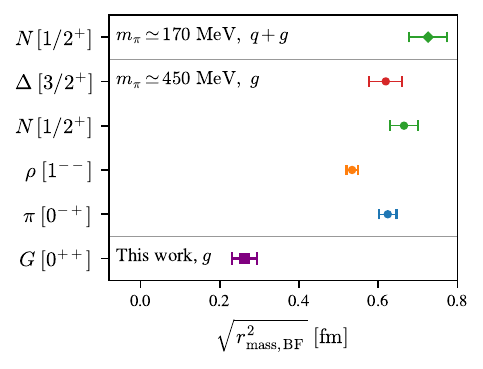}
\caption{The total (purely gluonic) mass radius of the glueball in the Breit frame obtained in this work, in units of $\mathrm{fm}$, compared with the gluon contribution to the BF mass radius of the pion, $\rho$-meson, nucleon, and $\Delta$-baryon extracted in a lattice QCD calculation with a heavier than physical pion mass $m_{\pi}\simeq 450~\text{MeV}$, as well as the \textit{total} mass radius of the nucleon extracted at $m_{\pi}\simeq 170~\text{MeV}$~\cite{Hackett:2023nkr,Hackett:2023rif}.
}
\label{fig:radii}
\end{figure}

The matrix elements defined in Eq.~\eqref{eq:ME} are obtained by using GEVP-optimized interpolating operators $\chi_0$ to compute vacuum-subtracted 
three-point functions~\cite{Blossier:2009kd} 
\begin{equation}\begin{split}
    C^{3\mathrm{pt}}_{0^{++},\mathcal{R}\ell}(\vec{p'}&,\vec{\Delta},t_s,\tau) 
    = \\
    \frac{1}{T} \sum_{t_0} &\braket{ \chi_0(\vec{p}', t_s+t_0) \, T_{\mathcal{R} \ell}(\vec{\Delta}, \tau+t_0) \, \chi_0(\vec{p}', t_0)^\dagger }
\end{split}\end{equation}
for $\left|\vec{\Delta}\right|^2 \leq 10 (2\pi/L)^2$ and $\left|\vec{p}'\right|^2 \leq 6 (2\pi/L)^2$. $T_{\mathcal{R} \ell}$ denotes the vacuum-subtracted gluon EMT projected to the $\ell$ row of the hypercubic group irreps~\cite{Gockeler:1996mu} $\mathcal{R}\in\{\tau_1^{(3)},\tau_3^{(6)}\}$. The individual EMT components are computed using the clover definition of the gluon field strength tensor, constructed using gluon links that have been acted on with 2 or 3 steps of stout smearing~\cite{Morningstar:2003gk}. The two- and three-point correlation functions are then combined into ratios:
\begin{equation}
\begin{split}
    R_{0^{++},\mathcal{R}\ell}(&\vec{p'},\vec{\Delta},t_s,\tau)
    = \frac{ 
        C^{3\mathrm{pt}}_{0^{++},\mathcal{R}\ell}(\vec{p'},\vec{\Delta},t_s,\tau)
    }{ C^{2\mathrm{pt}}_{0^{++}}(\vec{p'}, t_s) } \\
   \times&\sqrt{\frac{ C^{2\mathrm{pt}}_{0^{++}}(\vec{p}, t_s-\tau) }{ C^{2\mathrm{pt}}_{0^{++}}(\vec{p}', t_s-\tau)}\frac{ C^{2\mathrm{pt}}_{0^{++}}(\vec{p}', t_s)}
{C^{2\mathrm{pt}}_{0^{++}}(\vec{p}, t_s)}
\frac{C^{2\mathrm{pt}}_{0^{++}}(\vec{p}', \tau)}{C^{2\mathrm{pt}}_{0^{++}}(\vec{p}, \tau)}} \\
& \xrightarrow[\tau,t_s-\tau\gg 0]{} \frac{\left \langle G[0^{++}](p') \right| T^{\mu\nu} \left| G[0^{++}](p) \right \rangle}{2\sqrt{E_{\vec{p}}E_{\vec{p'}}}} .
\end{split} \label{eq:ratios}
\end{equation}
Ratios with choices of momenta, $\mathcal{R}$, and $\ell$ that result in the same linear combination of GFFs as defined in Eq.~\eqref{eq:ME} are averaged. 
These averaged ratios are fit to extract the ground-state contribution.
The fitting procedure is cross-checked against recent generalizations of the Lanczos method~\cite{Wagman:2024rid,Hackett:2024xnx,Hackett:2024nbe,Abbott:2025yhm}, which provides statistically compatible results.
Additional details, which closely follow previous lattice QCD studies of GFFs, e.g., Refs.~\cite{Brommel:2007zz,Detmold:2017oqb,Shanahan:2018pib,Shanahan:2018nnv,Hackett:2023rif,Hackett:2023nkr}, and figures showing the results of each analysis step, are given in the SM. 

Finally, the GFFs $A(t)$ and $D(t)$ are obtained by first grouping the fit results for the averaged ratios $\bar{R}_{0^{++},\mathcal{R}\ell}$ for each irrep $\mathcal{R}$ separately into 14 bins using k-means clustering \cite{kmeans1d} on the momentum transfer squared $t=\Delta^2$, then solving the overconstrained systems of linear equations dictated by Eq.~\eqref{eq:ME} to obtain the bare GFFs $A_\mathcal{R}(t)$ and $D_\mathcal{R}(t)$ for each irrep and bin. They can be renormalized by imposing the sum rule $A(0)=1$. The renormalization factors $1 / A^{\mathrm{bare}}_{\mathcal{R}}(0)$ are obtained from a fit of a $n$-pole model $\alpha / (1-t/\Lambda^2)^n$ to the bare GFF $A_{\mathcal{R}}(t)$; $\alpha$ and $\Lambda$ are fitted parameters, and $\alpha = A^{\mathrm{bare}}_{\mathcal{R}}(0)$. The bare GFFs in each momentum bin for each irrep are then multiplied by the renormalization factors, and then averaged together. Subsequently, the final results are obtained from a fit to the $n$-pole form, with $\alpha$ set to $1$ to enforce $A(0)=1$. Various analysis choices do not affect the results, as shown in the SM. Specifically, the choices $n\in\{1,2,3\}$ yield consistent values for the GFFs; in the main text $n=3$. Similarly, separate analyses using the bare three-point functions with 2 and 3 steps of stout smearing are consistent; in the main text 2 steps of stout smearing are used, as the $n$-pole fits to this data have a significantly larger p-value.

\textit{Results}: 
Figure~\ref{fig:GFFs} shows $A(t)$ and $D(t)$ of the $0^{++}$ glueball determined in this work.
Also shown for comparison are
the gluon GFFs, up to a factor\footnote{Mesons and baryons with valence quarks and antiquarks also receive a quark contribution to their GFFs, which mixes with the gluonic component. The work of Ref.~\cite{Pefkou:2021fni} constrained only the gluon contribution of the GFFs, also neglecting the mixing. The comparison of the overall normalization between those results and the gluon GFFs in this work---which coincide with the total GFFs in a theory with only gluonic degrees of freedom, as investigated here---is not meaningful.} of $A(0)$, of four hadrons with quantum numbers $J^P=0^{-}$, $1^{-}$, $1/2^+$, and $3/2^+$, corresponding to the pion, $\rho$ meson, nucleon, and $\Delta$ baryon, computed using a 
single lattice QCD ensemble with $N_f=2+1$ clover-improved dynamical quark flavors, $a \simeq 0.12~\text{fm}$, and $m_{\pi} \simeq 450~\text{MeV}$~\cite{Pefkou:2021fni}. While the present calculation is undertaken in Yang-Mills theory, this qualitative comparison may nevertheless yield interesting conclusions; a quenched calculation of the gluon GFFs of a pseudoscalar meson state with valence quark masses tuned such that the hadron mass matches that of Ref.~\cite{Pefkou:2021fni} (shown in the SM) is consistent with the QCD results for the pion shown in Fig.~\ref{fig:GFFs}, indicating that quenching has a negligible effect on these quantities at the level of uncertainties of this calculation. The glueball $A(t)$ form factor decays more slowly than that of the pion, corresponding to a smaller mass radius contribution. The same is observed in the respective $D$ form factor, although the statistical uncertainty of $D(t)$ is larger for the glueball\footnote{The flat $t$-dependence of $D(t)$ is difficult to interpret at this level of statistics, especially when taking into consideration the 3-stout smeared result shown in the SM, where the behavior is more consistent with monotonically increasing.}.

Fig.~\ref{fig:170} shows a comparison of the results of this work with the \textit{total} GFFs of the pion~\cite{Hackett:2023nkr} and the nucleon~\cite{Hackett:2023rif}, computed using an ensemble with the same action as in Ref.~\cite{Pefkou:2021fni} but smaller lattice spacing $a\simeq 0.091~\mathrm{fm}$ and closer to the physical point with $m_{\pi}\simeq 170~\text{MeV}$. Since this comparison shows {\it total} GFFs, no division by $A(0)$ is necessary. The same qualitative conclusions are drawn; the glueball GFFs decay more slowly than those of the other hadrons, implying a smaller radius.

To carefully assess the radii comparison, we consider the root mean square radius of the energy density of the glueball in the Breit frame, also known as the mass radius. It is~\cite{Polyakov:2018zvc,Freese:2019bhb} 
\begin{equation}
r^2_{\mathrm{mass,BF}} = \left. \frac{1}{A(0)}\left[6\frac{dA(t)}{dt}\right\rvert_{t=0}-\frac{3}{4m^2}(A(0)
+2D(0))\right] ,
\end{equation}
and is found to be $0.263(31)~\text{fm}$. This is remarkably small, and close to older lattice QCD and Bethe-Salpeter amplitude findings, which estimated the size of the scalar glueball to be in the range $0.1-0.3~\text{fm}$~\cite{PhysRevD.43.2301,PhysRevLett.69.245,Loan:2006gm,Schafer:1994fd,Forkel:2000fd} by analyzing its (Bethe-Salpeter) wavefunction. This supports the view that the scalar glueball size is set by short-range gluon interactions (e.g.~instanton-induced forces~\cite{Schafer:1994fd,Forkel:2000fd}) and not the confinement scale of QCD. It is also smaller than a dispersive prediction of the scalar radius of the $\sigma$ meson~\cite{Albaladejo:2012te}, suggesting that the scalar glueball is the most compact state in that channel.

Figure~\ref{fig:radii} shows a comparison of this quantity with the results of the $n$-pole fits of Refs.~\cite{Pefkou:2021fni,Hackett:2023rif} for the other hadron states\footnote{The numbers in Fig.~\ref{fig:radii} taken from Ref.~\cite{Pefkou:2021fni} differ from those currently in Table~X of the published article. This is due to an error in normalization in the results of that table only; an erratum is forthcoming.}\footnote{We do not compare with the radius of the $m_{\pi} \simeq 170$~MeV pion study of Ref.~\cite{Hackett:2023nkr}, because for light hadrons the 3D BF density interpretation is expected to break down~\cite{Yennie:1957skg,Burkardt:2002hr,Miller:2018ybm,Jaffe:2020ebz,Epelbaum:2022fjc,Freese:2021czn,Lorce:2018egm,Freese:2025tqd}.}.  Note that other than for the pion, the BF mass radius of the rest of the hadrons depends on additional GFFs besides $A(t)$ and $D(t)$~\cite{Polyakov:2018zvc,Polyakov:2019lbq,Sun:2020wfo,Kim:2020lrs}, and their contribution is included in the radii values of Fig.~\ref{fig:radii}\footnote{This excludes the $D_{1,g}^{\rho}$ form factor for $\rho$ and the $F_{11}^{\Delta,g}$, $F_{21}^{\Delta,g}$, $F_{41}^{\Delta,g}$, and $F_{51}^{\Delta,g}$ form factors for $\Delta$ that were consistent with zero and were not fit with a tripole form in Ref.~\cite{Pefkou:2021fni}.}. The scalar glueball radius is considerably smaller than that of the other hadrons.

\textit{Conclusion}: 
These results constitute the first investigation of the internal structure of glueballs
in an SU(3) lattice gauge theory, marking a promising advance in the understanding of the internal structure of potential glueball-like hadrons in nature.

While the calculation is undertaken in Yang-Mills theory instead of QCD, comparison with (quenched and full) QCD calculations of the (gluon and total) GFFs of other hadrons provides the first
indication that glueballs may have a different gluonic structure than more typical hadronic states; specifically, the $A(t)$ GFF decays more slowly, corresponding to a smaller gluonic radius. This observation, if it is maintained in future QCD calculations of glueball GFFs, would open the potential for future smoking gun identification of glueball states directly via their gluon structure. 

It would also be interesting to repeat this study with other glueball states in Yang-Mills theory, for instance with pseudoscalar or tensor quantum numbers. Notably, early lattice QCD studies based on wavefunction analyses suggested that the tensor glueball has a significantly larger radius than the scalar~\cite{PhysRevD.43.2301,PhysRevLett.69.245,Loan:2006gm}. Future calculations could benefit from recent multi-level algorithm development~\cite{Barca:2024fpc,Barca:2025fdq}.

\textit{Acknowledgments:} The authors acknowledge the MIT SuperCloud~\cite{reuther2018interactive} and Lincoln Laboratory Supercomputing Center for providing HPC resources that have contributed to the research results reported within this paper. This work made use of resources provided by subMIT at MIT Physics~\cite{bendavid2025submitphysicsanalysisfacility}.

The authors thank Julian Urban for contributions at early stages of the project. We also thank Lorenzo Barca, Max Hansen, and André Walker-Loud for useful discussions. PES is supported in part by the U.S.~Department of Energy, Office of Science, Office of Nuclear Physics, under grant Contract Number DE-SC0011090, by Early Career Award DE-SC0021006, by Simons Foundation grant 994314 (Simons Collaboration on Confinement and QCD Strings), by the U.S. Department of Energy SciDAC5 award DE-SC0023116, and has benefited from the QGT Topical Collaboration DE-SC0023646.
DAP is supported from the Office of Nuclear Physics, Department of Energy, under contract DE-SC0004658.
This document was prepared using the resources of the Fermi National Accelerator Laboratory (Fermilab), a U.S. Department of Energy, Office of Science, Office of High Energy Physics HEP User Facility. Fermilab is managed by Fermi Forward Discovery Group, LLC, acting under Contract No.~89243024CSC000002. DAP thanks the Albert Einstein Center at the University of Bern for its hospitality during a visit that significantly advanced this work.
This research used resources of the National Energy Research Scientific Computing Center (NERSC), a U.S. Department of Energy Office of Science User Facility operated under Contract No.~DE-AC02-05CH11231. 
RA is supported by the U.S. Department of Energy SciDAC5 award DE-SC0023116 and the High Energy Physics Computing Traineeship for Lattice Gauge Theory (DE-SC0024053).
PES thanks the Institute for Nuclear Theory at the University of Washington for its kind hospitality and stimulating research environment. This research was supported in part by the INT's U.S. Department of Energy grant No. DE-FG02-00ER41132. This manuscript was finalized at Aspen Center for Physics, which is supported by National Science Foundation grant PHY-2210452.

The Chroma~\cite{Edwards:2004sx} and  LALIBE~\cite{lalibe} and JAX~\cite{jax2018github} libraries were used in this work.
Data analysis used NumPy~\cite{harris2020array}, SciPy~\cite{2020SciPy-NMeth}, lsqfit~\cite{peter_lepage_2020_4037174}, and gvar~\cite{peter_lepage_2020_4290884}.
Figures were produced using matplotlib~\cite{Hunter:2007}.

\bibliography{main}

\begin{thebibliography}{90}%
\makeatletter
\providecommand \@ifxundefined [1]{%
 \@ifx{#1\undefined}
}%
\providecommand \@ifnum [1]{%
 \ifnum #1\expandafter \@firstoftwo
 \else \expandafter \@secondoftwo
 \fi
}%
\providecommand \@ifx [1]{%
 \ifx #1\expandafter \@firstoftwo
 \else \expandafter \@secondoftwo
 \fi
}%
\providecommand \natexlab [1]{#1}%
\providecommand \enquote  [1]{``#1''}%
\providecommand \bibnamefont  [1]{#1}%
\providecommand \bibfnamefont [1]{#1}%
\providecommand \citenamefont [1]{#1}%
\providecommand \href@noop [0]{\@secondoftwo}%
\providecommand \href [0]{\begingroup \@sanitize@url \@href}%
\providecommand \@href[1]{\@@startlink{#1}\@@href}%
\providecommand \@@href[1]{\endgroup#1\@@endlink}%
\providecommand \@sanitize@url [0]{\catcode `\\12\catcode `\$12\catcode `\&12\catcode `\#12\catcode `\^12\catcode `\_12\catcode `\%12\relax}%
\providecommand \@@startlink[1]{}%
\providecommand \@@endlink[0]{}%
\providecommand \url  [0]{\begingroup\@sanitize@url \@url }%
\providecommand \@url [1]{\endgroup\@href {#1}{\urlprefix }}%
\providecommand \urlprefix  [0]{URL }%
\providecommand \Eprint [0]{\href }%
\providecommand \doibase [0]{https://doi.org/}%
\providecommand \selectlanguage [0]{\@gobble}%
\providecommand \bibinfo  [0]{\@secondoftwo}%
\providecommand \bibfield  [0]{\@secondoftwo}%
\providecommand \translation [1]{[#1]}%
\providecommand \BibitemOpen [0]{}%
\providecommand \bibitemStop [0]{}%
\providecommand \bibitemNoStop [0]{.\EOS\space}%
\providecommand \EOS [0]{\spacefactor3000\relax}%
\providecommand \BibitemShut  [1]{\csname bibitem#1\endcsname}%
\let\auto@bib@innerbib\@empty
%</preamble>
\bibitem [{\citenamefont {Fritzsch}\ and\ \citenamefont {Gell-Mann}(1972)}]{Fritzsch:1972jv}%
  \BibitemOpen
  \bibfield  {author} {\bibinfo {author} {\bibfnamefont {H.}~\bibnamefont {Fritzsch}}\ and\ \bibinfo {author} {\bibfnamefont {M.}~\bibnamefont {Gell-Mann}},\ }\bibfield  {title} {\bibinfo {title} {{Current algebra: Quarks and what else?}},\ }\href@noop {} {\bibfield  {journal} {\bibinfo  {journal} {eConf}\ }\textbf {\bibinfo {volume} {C720906V2}},\ \bibinfo {pages} {135} (\bibinfo {year} {1972})},\ \Eprint {https://arxiv.org/abs/hep-ph/0208010} {arXiv:hep-ph/0208010} \BibitemShut {NoStop}%
\bibitem [{\citenamefont {Klempt}\ and\ \citenamefont {Zaitsev}(2007)}]{Klempt:2007cp}%
  \BibitemOpen
  \bibfield  {author} {\bibinfo {author} {\bibfnamefont {E.}~\bibnamefont {Klempt}}\ and\ \bibinfo {author} {\bibfnamefont {A.}~\bibnamefont {Zaitsev}},\ }\bibfield  {title} {\bibinfo {title} {{Glueballs, Hybrids, Multiquarks. Experimental facts versus QCD inspired concepts}},\ }\href {https://doi.org/10.1016/j.physrep.2007.07.006} {\bibfield  {journal} {\bibinfo  {journal} {Phys. Rept.}\ }\textbf {\bibinfo {volume} {454}},\ \bibinfo {pages} {1} (\bibinfo {year} {2007})},\ \Eprint {https://arxiv.org/abs/0708.4016} {arXiv:0708.4016 [hep-ph]} \BibitemShut {NoStop}%
\bibitem [{\citenamefont {Crede}\ and\ \citenamefont {Meyer}(2009)}]{Crede:2008vw}%
  \BibitemOpen
  \bibfield  {author} {\bibinfo {author} {\bibfnamefont {V.}~\bibnamefont {Crede}}\ and\ \bibinfo {author} {\bibfnamefont {C.~A.}\ \bibnamefont {Meyer}},\ }\bibfield  {title} {\bibinfo {title} {{The Experimental Status of Glueballs}},\ }\href {https://doi.org/10.1016/j.ppnp.2009.03.001} {\bibfield  {journal} {\bibinfo  {journal} {Prog. Part. Nucl. Phys.}\ }\textbf {\bibinfo {volume} {63}},\ \bibinfo {pages} {74} (\bibinfo {year} {2009})},\ \Eprint {https://arxiv.org/abs/0812.0600} {arXiv:0812.0600 [hep-ex]} \BibitemShut {NoStop}%
\bibitem [{\citenamefont {Chen}\ \emph {et~al.}(2023)\citenamefont {Chen}, \citenamefont {Chen}, \citenamefont {Liu}, \citenamefont {Liu},\ and\ \citenamefont {Zhu}}]{Chen:2022asf}%
  \BibitemOpen
  \bibfield  {author} {\bibinfo {author} {\bibfnamefont {H.-X.}\ \bibnamefont {Chen}}, \bibinfo {author} {\bibfnamefont {W.}~\bibnamefont {Chen}}, \bibinfo {author} {\bibfnamefont {X.}~\bibnamefont {Liu}}, \bibinfo {author} {\bibfnamefont {Y.-R.}\ \bibnamefont {Liu}},\ and\ \bibinfo {author} {\bibfnamefont {S.-L.}\ \bibnamefont {Zhu}},\ }\bibfield  {title} {\bibinfo {title} {{An updated review of the new hadron states}},\ }\href {https://doi.org/10.1088/1361-6633/aca3b6} {\bibfield  {journal} {\bibinfo  {journal} {Rept. Prog. Phys.}\ }\textbf {\bibinfo {volume} {86}},\ \bibinfo {pages} {026201} (\bibinfo {year} {2023})},\ \Eprint {https://arxiv.org/abs/2204.02649} {arXiv:2204.02649 [hep-ph]} \BibitemShut {NoStop}%
\bibitem [{\citenamefont {Ablikim}\ \emph {et~al.}(2011)\citenamefont {Ablikim} \emph {et~al.}}]{BESIII:2010gmv}%
  \BibitemOpen
  \bibfield  {author} {\bibinfo {author} {\bibfnamefont {M.}~\bibnamefont {Ablikim}} \emph {et~al.} (\bibinfo {collaboration} {BESIII}),\ }\bibfield  {title} {\bibinfo {title} {{Confirmation of the $X(1835)$ and observation of the resonances $X(2120)$ and $X(2370)$ in $J/\psi\to \gamma \pi^+\pi^-\eta^\prime$}},\ }\href {https://doi.org/10.1103/PhysRevLett.106.072002} {\bibfield  {journal} {\bibinfo  {journal} {Phys. Rev. Lett.}\ }\textbf {\bibinfo {volume} {106}},\ \bibinfo {pages} {072002} (\bibinfo {year} {2011})},\ \Eprint {https://arxiv.org/abs/1012.3510} {arXiv:1012.3510 [hep-ex]} \BibitemShut {NoStop}%
\bibitem [{\citenamefont {Ablikim}\ \emph {et~al.}(2020)\citenamefont {Ablikim} \emph {et~al.}}]{BESIII:2019wkp}%
  \BibitemOpen
  \bibfield  {author} {\bibinfo {author} {\bibfnamefont {M.}~\bibnamefont {Ablikim}} \emph {et~al.} (\bibinfo {collaboration} {BESIII}),\ }\bibfield  {title} {\bibinfo {title} {{Observation of $X(2370)$ and search for X(2120) in $J/\psi \rightarrow \gamma K{\bar{K}} \eta '$}},\ }\href {https://doi.org/10.1140/epjc/s10052-020-8078-4} {\bibfield  {journal} {\bibinfo  {journal} {Eur. Phys. J. C}\ }\textbf {\bibinfo {volume} {80}},\ \bibinfo {pages} {746} (\bibinfo {year} {2020})},\ \Eprint {https://arxiv.org/abs/1912.11253} {arXiv:1912.11253 [hep-ex]} \BibitemShut {NoStop}%
\bibitem [{\citenamefont {Ablikim}\ \emph {et~al.}(2024)\citenamefont {Ablikim} \emph {et~al.}}]{BESIII:2023wfi}%
  \BibitemOpen
  \bibfield  {author} {\bibinfo {author} {\bibfnamefont {M.}~\bibnamefont {Ablikim}} \emph {et~al.} (\bibinfo {collaboration} {BESIII}),\ }\bibfield  {title} {\bibinfo {title} {{Determination of Spin-Parity Quantum Numbers of X(2370) as 0-+ from J/\ensuremath{\psi}\textrightarrow{}\ensuremath{\gamma}KS0KS0\ensuremath{\eta}'}},\ }\href {https://doi.org/10.1103/PhysRevLett.132.181901} {\bibfield  {journal} {\bibinfo  {journal} {Phys. Rev. Lett.}\ }\textbf {\bibinfo {volume} {132}},\ \bibinfo {pages} {181901} (\bibinfo {year} {2024})},\ \Eprint {https://arxiv.org/abs/2312.05324} {arXiv:2312.05324 [hep-ex]} \BibitemShut {NoStop}%
\bibitem [{\citenamefont {Teper}(1998)}]{Teper:1998kw}%
  \BibitemOpen
  \bibfield  {author} {\bibinfo {author} {\bibfnamefont {M.~J.}\ \bibnamefont {Teper}},\ }\bibfield  {title} {\bibinfo {title} {{Glueball masses and other physical properties of SU(N) gauge theories in D = (3+1): A Review of lattice results for theorists}},\ }\href@noop {} {\  (\bibinfo {year} {1998})},\ \Eprint {https://arxiv.org/abs/hep-th/9812187} {arXiv:hep-th/9812187} \BibitemShut {NoStop}%
\bibitem [{\citenamefont {Morningstar}\ and\ \citenamefont {Peardon}(1999)}]{Morningstar:1999rf}%
  \BibitemOpen
  \bibfield  {author} {\bibinfo {author} {\bibfnamefont {C.~J.}\ \bibnamefont {Morningstar}}\ and\ \bibinfo {author} {\bibfnamefont {M.~J.}\ \bibnamefont {Peardon}},\ }\bibfield  {title} {\bibinfo {title} {{The Glueball spectrum from an anisotropic lattice study}},\ }\href {https://doi.org/10.1103/PhysRevD.60.034509} {\bibfield  {journal} {\bibinfo  {journal} {Phys. Rev. D}\ }\textbf {\bibinfo {volume} {60}},\ \bibinfo {pages} {034509} (\bibinfo {year} {1999})},\ \Eprint {https://arxiv.org/abs/hep-lat/9901004} {arXiv:hep-lat/9901004} \BibitemShut {NoStop}%
\bibitem [{\citenamefont {Athenodorou}\ and\ \citenamefont {Teper}(2017)}]{Athenodorou:2016ebg}%
  \BibitemOpen
  \bibfield  {author} {\bibinfo {author} {\bibfnamefont {A.}~\bibnamefont {Athenodorou}}\ and\ \bibinfo {author} {\bibfnamefont {M.}~\bibnamefont {Teper}},\ }\bibfield  {title} {\bibinfo {title} {{SU(N) gauge theories in 2+1 dimensions: glueball spectra and k-string tensions}},\ }\href {https://doi.org/10.1007/JHEP02(2017)015} {\bibfield  {journal} {\bibinfo  {journal} {JHEP}\ }\textbf {\bibinfo {volume} {02}},\ \bibinfo {pages} {015}},\ \Eprint {https://arxiv.org/abs/1609.03873} {arXiv:1609.03873 [hep-lat]} \BibitemShut {NoStop}%
\bibitem [{\citenamefont {Athenodorou}\ and\ \citenamefont {Teper}(2020)}]{Athenodorou:2020ani}%
  \BibitemOpen
  \bibfield  {author} {\bibinfo {author} {\bibfnamefont {A.}~\bibnamefont {Athenodorou}}\ and\ \bibinfo {author} {\bibfnamefont {M.}~\bibnamefont {Teper}},\ }\bibfield  {title} {\bibinfo {title} {{The glueball spectrum of SU(3) gauge theory in 3 + 1 dimensions}},\ }\href {https://doi.org/10.1007/JHEP11(2020)172} {\bibfield  {journal} {\bibinfo  {journal} {JHEP}\ }\textbf {\bibinfo {volume} {11}},\ \bibinfo {pages} {172}},\ \Eprint {https://arxiv.org/abs/2007.06422} {arXiv:2007.06422 [hep-lat]} \BibitemShut {NoStop}%
\bibitem [{\citenamefont {Sakai}\ and\ \citenamefont {Sasaki}(2023)}]{Sakai:2022zdc}%
  \BibitemOpen
  \bibfield  {author} {\bibinfo {author} {\bibfnamefont {K.}~\bibnamefont {Sakai}}\ and\ \bibinfo {author} {\bibfnamefont {S.}~\bibnamefont {Sasaki}},\ }\bibfield  {title} {\bibinfo {title} {{Glueball spectroscopy in lattice QCD using gradient flow}},\ }\href {https://doi.org/10.1103/PhysRevD.107.034510} {\bibfield  {journal} {\bibinfo  {journal} {Phys. Rev. D}\ }\textbf {\bibinfo {volume} {107}},\ \bibinfo {pages} {034510} (\bibinfo {year} {2023})},\ \Eprint {https://arxiv.org/abs/2211.15176} {arXiv:2211.15176 [hep-lat]} \BibitemShut {NoStop}%
\bibitem [{\citenamefont {Vadacchino}(2023)}]{Vadacchino:2023vnc}%
  \BibitemOpen
  \bibfield  {author} {\bibinfo {author} {\bibfnamefont {D.}~\bibnamefont {Vadacchino}},\ }\bibfield  {title} {\bibinfo {title} {{A review on Glueball hunting}},\ }in\ \href@noop {} {\emph {\bibinfo {booktitle} {{39th International Symposium on Lattice Field Theory}}}}\ (\bibinfo {year} {2023})\ \Eprint {https://arxiv.org/abs/2305.04869} {arXiv:2305.04869 [hep-lat]} \BibitemShut {NoStop}%
\bibitem [{\citenamefont {Morningstar}(2024)}]{Morningstar:2024vjk}%
  \BibitemOpen
  \bibfield  {author} {\bibinfo {author} {\bibfnamefont {C.}~\bibnamefont {Morningstar}},\ }\bibfield  {title} {\bibinfo {title} {{Update on Glueballs}},\ }\href {https://doi.org/10.22323/1.466.0004} {\bibfield  {journal} {\bibinfo  {journal} {PoS}\ }\textbf {\bibinfo {volume} {LATTICE2024}},\ \bibinfo {pages} {004} (\bibinfo {year} {2024})},\ \Eprint {https://arxiv.org/abs/2502.02547} {arXiv:2502.02547 [hep-lat]} \BibitemShut {NoStop}%
\bibitem [{\citenamefont {Bali}\ \emph {et~al.}(2000)\citenamefont {Bali}, \citenamefont {Bolder}, \citenamefont {Eicker}, \citenamefont {Lippert}, \citenamefont {Orth}, \citenamefont {Ueberholz}, \citenamefont {Schilling},\ and\ \citenamefont {Struckmann}}]{Bali:2000vr}%
  \BibitemOpen
  \bibfield  {author} {\bibinfo {author} {\bibfnamefont {G.~S.}\ \bibnamefont {Bali}}, \bibinfo {author} {\bibfnamefont {B.}~\bibnamefont {Bolder}}, \bibinfo {author} {\bibfnamefont {N.}~\bibnamefont {Eicker}}, \bibinfo {author} {\bibfnamefont {T.}~\bibnamefont {Lippert}}, \bibinfo {author} {\bibfnamefont {B.}~\bibnamefont {Orth}}, \bibinfo {author} {\bibfnamefont {P.}~\bibnamefont {Ueberholz}}, \bibinfo {author} {\bibfnamefont {K.}~\bibnamefont {Schilling}},\ and\ \bibinfo {author} {\bibfnamefont {T.}~\bibnamefont {Struckmann}} (\bibinfo {collaboration} {TXL, T(X)L}),\ }\bibfield  {title} {\bibinfo {title} {{Static potentials and glueball masses from QCD simulations with Wilson sea quarks}},\ }\href {https://doi.org/10.1103/PhysRevD.62.054503} {\bibfield  {journal} {\bibinfo  {journal} {Phys. Rev. D}\ }\textbf {\bibinfo {volume} {62}},\ \bibinfo {pages} {054503} (\bibinfo {year} {2000})},\ \Eprint {https://arxiv.org/abs/hep-lat/0003012} {arXiv:hep-lat/0003012} \BibitemShut {NoStop}%
\bibitem [{\citenamefont {Athenodorou}\ \emph {et~al.}(2023)\citenamefont {Athenodorou}, \citenamefont {Finkenrath}, \citenamefont {Lantos},\ and\ \citenamefont {Teper}}]{Athenodorou:2023ntf}%
  \BibitemOpen
  \bibfield  {author} {\bibinfo {author} {\bibfnamefont {A.}~\bibnamefont {Athenodorou}}, \bibinfo {author} {\bibfnamefont {J.}~\bibnamefont {Finkenrath}}, \bibinfo {author} {\bibfnamefont {A.}~\bibnamefont {Lantos}},\ and\ \bibinfo {author} {\bibfnamefont {M.}~\bibnamefont {Teper}},\ }\bibfield  {title} {\bibinfo {title} {{Glueball Spectrum with four light dynamical fermions}},\ }\href@noop {} {\  (\bibinfo {year} {2023})},\ \Eprint {https://arxiv.org/abs/2308.10054} {arXiv:2308.10054 [hep-lat]} \BibitemShut {NoStop}%
\bibitem [{\citenamefont {Sch{\"a}fer}\ and\ \citenamefont {Shuryak}(1995)}]{Schafer:1994fd}%
  \BibitemOpen
  \bibfield  {author} {\bibinfo {author} {\bibfnamefont {T.}~\bibnamefont {Sch{\"a}fer}}\ and\ \bibinfo {author} {\bibfnamefont {E.~V.}\ \bibnamefont {Shuryak}},\ }\bibfield  {title} {\bibinfo {title} {{Glueballs and instantons}},\ }\href {https://doi.org/10.1103/PhysRevLett.75.1707} {\bibfield  {journal} {\bibinfo  {journal} {Phys. Rev. Lett.}\ }\textbf {\bibinfo {volume} {75}},\ \bibinfo {pages} {1707} (\bibinfo {year} {1995})},\ \Eprint {https://arxiv.org/abs/hep-ph/9410372} {arXiv:hep-ph/9410372} \BibitemShut {NoStop}%
\bibitem [{\citenamefont {Forkel}(2001)}]{Forkel:2000fd}%
  \BibitemOpen
  \bibfield  {author} {\bibinfo {author} {\bibfnamefont {H.}~\bibnamefont {Forkel}},\ }\bibfield  {title} {\bibinfo {title} {{Scalar gluonium and instantons}},\ }\href {https://doi.org/10.1103/PhysRevD.64.034015} {\bibfield  {journal} {\bibinfo  {journal} {Phys. Rev. D}\ }\textbf {\bibinfo {volume} {64}},\ \bibinfo {pages} {034015} (\bibinfo {year} {2001})},\ \Eprint {https://arxiv.org/abs/hep-ph/0005004} {arXiv:hep-ph/0005004} \BibitemShut {NoStop}%
\bibitem [{\citenamefont {Hou}\ and\ \citenamefont {Wong}(2003)}]{Hou:2002jv}%
  \BibitemOpen
  \bibfield  {author} {\bibinfo {author} {\bibfnamefont {W.-S.}\ \bibnamefont {Hou}}\ and\ \bibinfo {author} {\bibfnamefont {G.-G.}\ \bibnamefont {Wong}},\ }\bibfield  {title} {\bibinfo {title} {{The Glueball spectrum from a potential model}},\ }\href {https://doi.org/10.1103/PhysRevD.67.034003} {\bibfield  {journal} {\bibinfo  {journal} {Phys. Rev. D}\ }\textbf {\bibinfo {volume} {67}},\ \bibinfo {pages} {034003} (\bibinfo {year} {2003})},\ \Eprint {https://arxiv.org/abs/hep-ph/0207292} {arXiv:hep-ph/0207292} \BibitemShut {NoStop}%
\bibitem [{\citenamefont {Gupta}\ \emph {et~al.}(1991)\citenamefont {Gupta}, \citenamefont {Patel}, \citenamefont {Baillie}, \citenamefont {Kilcup},\ and\ \citenamefont {Sharpe}}]{PhysRevD.43.2301}%
  \BibitemOpen
  \bibfield  {author} {\bibinfo {author} {\bibfnamefont {R.}~\bibnamefont {Gupta}}, \bibinfo {author} {\bibfnamefont {A.}~\bibnamefont {Patel}}, \bibinfo {author} {\bibfnamefont {C.~F.}\ \bibnamefont {Baillie}}, \bibinfo {author} {\bibfnamefont {G.~W.}\ \bibnamefont {Kilcup}},\ and\ \bibinfo {author} {\bibfnamefont {S.~R.}\ \bibnamefont {Sharpe}},\ }\bibfield  {title} {\bibinfo {title} {Exploring glueball wave functions on the lattice},\ }\href {https://doi.org/10.1103/PhysRevD.43.2301} {\bibfield  {journal} {\bibinfo  {journal} {Phys. Rev. D}\ }\textbf {\bibinfo {volume} {43}},\ \bibinfo {pages} {2301} (\bibinfo {year} {1991})}\BibitemShut {NoStop}%
\bibitem [{\citenamefont {de~Forcrand}\ and\ \citenamefont {Liu}(1992)}]{PhysRevLett.69.245}%
  \BibitemOpen
  \bibfield  {author} {\bibinfo {author} {\bibfnamefont {P.}~\bibnamefont {de~Forcrand}}\ and\ \bibinfo {author} {\bibfnamefont {K.-F.}\ \bibnamefont {Liu}},\ }\bibfield  {title} {\bibinfo {title} {Glueball wave functions in lattice gauge calculations},\ }\href {https://doi.org/10.1103/PhysRevLett.69.245} {\bibfield  {journal} {\bibinfo  {journal} {Phys. Rev. Lett.}\ }\textbf {\bibinfo {volume} {69}},\ \bibinfo {pages} {245} (\bibinfo {year} {1992})}\BibitemShut {NoStop}%
\bibitem [{\citenamefont {Loan}\ and\ \citenamefont {Ying}(2006)}]{Loan:2006gm}%
  \BibitemOpen
  \bibfield  {author} {\bibinfo {author} {\bibfnamefont {M.}~\bibnamefont {Loan}}\ and\ \bibinfo {author} {\bibfnamefont {Y.}~\bibnamefont {Ying}},\ }\bibfield  {title} {\bibinfo {title} {{Sizes of lightest glueballs in SU(3) lattice gauge theory}},\ }\href {https://doi.org/10.1143/PTP.116.169} {\bibfield  {journal} {\bibinfo  {journal} {Prog. Theor. Phys.}\ }\textbf {\bibinfo {volume} {116}},\ \bibinfo {pages} {169} (\bibinfo {year} {2006})},\ \Eprint {https://arxiv.org/abs/hep-lat/0603030} {arXiv:hep-lat/0603030} \BibitemShut {NoStop}%
\bibitem [{\citenamefont {Tickle}\ and\ \citenamefont {Michael}(1990)}]{Tickle:1989gw}%
  \BibitemOpen
  \bibfield  {author} {\bibinfo {author} {\bibfnamefont {G.~A.}\ \bibnamefont {Tickle}}\ and\ \bibinfo {author} {\bibfnamefont {C.}~\bibnamefont {Michael}},\ }\bibfield  {title} {\bibinfo {title} {{An Investigation of the Structure of the O+ Glueball in SU(2) Lattice Gauge Theory}},\ }\href {https://doi.org/10.1016/0550-3213(90)90053-G} {\bibfield  {journal} {\bibinfo  {journal} {Nucl. Phys. B}\ }\textbf {\bibinfo {volume} {333}},\ \bibinfo {pages} {593} (\bibinfo {year} {1990})}\BibitemShut {NoStop}%
\bibitem [{\citenamefont {Albaladejo}\ and\ \citenamefont {Oller}(2012)}]{Albaladejo:2012te}%
  \BibitemOpen
  \bibfield  {author} {\bibinfo {author} {\bibfnamefont {M.}~\bibnamefont {Albaladejo}}\ and\ \bibinfo {author} {\bibfnamefont {J.~A.}\ \bibnamefont {Oller}},\ }\bibfield  {title} {\bibinfo {title} {{On the size of the sigma meson and its nature}},\ }\href {https://doi.org/10.1103/PhysRevD.86.034003} {\bibfield  {journal} {\bibinfo  {journal} {Phys. Rev. D}\ }\textbf {\bibinfo {volume} {86}},\ \bibinfo {pages} {034003} (\bibinfo {year} {2012})},\ \Eprint {https://arxiv.org/abs/1205.6606} {arXiv:1205.6606 [hep-ph]} \BibitemShut {NoStop}%
\bibitem [{\citenamefont {Polyakov}(2003)}]{Polyakov:2002yz}%
  \BibitemOpen
  \bibfield  {author} {\bibinfo {author} {\bibfnamefont {M.}~\bibnamefont {Polyakov}},\ }\bibfield  {title} {\bibinfo {title} {{Generalized parton distributions and strong forces inside nucleons and nuclei}},\ }\href {https://doi.org/10.1016/S0370-2693(03)00036-4} {\bibfield  {journal} {\bibinfo  {journal} {Phys. Lett. B}\ }\textbf {\bibinfo {volume} {555}},\ \bibinfo {pages} {57} (\bibinfo {year} {2003})},\ \Eprint {https://arxiv.org/abs/hep-ph/0210165} {arXiv:hep-ph/0210165} \BibitemShut {NoStop}%
\bibitem [{\citenamefont {Lorc\'e}\ \emph {et~al.}(2019)\citenamefont {Lorc\'e}, \citenamefont {Moutarde},\ and\ \citenamefont {Trawi\'nski}}]{Lorce:2018egm}%
  \BibitemOpen
  \bibfield  {author} {\bibinfo {author} {\bibfnamefont {C.}~\bibnamefont {Lorc\'e}}, \bibinfo {author} {\bibfnamefont {H.}~\bibnamefont {Moutarde}},\ and\ \bibinfo {author} {\bibfnamefont {A.~P.}\ \bibnamefont {Trawi\'nski}},\ }\bibfield  {title} {\bibinfo {title} {{Revisiting the mechanical properties of the nucleon}},\ }\href {https://doi.org/10.1140/epjc/s10052-019-6572-3} {\bibfield  {journal} {\bibinfo  {journal} {Eur. Phys. J. C}\ }\textbf {\bibinfo {volume} {79}},\ \bibinfo {pages} {89} (\bibinfo {year} {2019})},\ \Eprint {https://arxiv.org/abs/1810.09837} {arXiv:1810.09837 [hep-ph]} \BibitemShut {NoStop}%
\bibitem [{\citenamefont {Polyakov}\ and\ \citenamefont {Schweitzer}(2018)}]{Polyakov:2018zvc}%
  \BibitemOpen
  \bibfield  {author} {\bibinfo {author} {\bibfnamefont {M.~V.}\ \bibnamefont {Polyakov}}\ and\ \bibinfo {author} {\bibfnamefont {P.}~\bibnamefont {Schweitzer}},\ }\bibfield  {title} {\bibinfo {title} {{Forces inside hadrons: pressure, surface tension, mechanical radius, and all that}},\ }\href {https://doi.org/10.1142/S0217751X18300259} {\bibfield  {journal} {\bibinfo  {journal} {Int. J. Mod. Phys. A}\ }\textbf {\bibinfo {volume} {33}},\ \bibinfo {pages} {1830025} (\bibinfo {year} {2018})},\ \Eprint {https://arxiv.org/abs/1805.06596} {arXiv:1805.06596 [hep-ph]} \BibitemShut {NoStop}%
\bibitem [{\citenamefont {Burkert}\ \emph {et~al.}(2023)\citenamefont {Burkert}, \citenamefont {Elouadrhiri}, \citenamefont {Girod}, \citenamefont {Lorc\'e}, \citenamefont {Schweitzer},\ and\ \citenamefont {Shanahan}}]{Burkert:2023wzr}%
  \BibitemOpen
  \bibfield  {author} {\bibinfo {author} {\bibfnamefont {V.~D.}\ \bibnamefont {Burkert}}, \bibinfo {author} {\bibfnamefont {L.}~\bibnamefont {Elouadrhiri}}, \bibinfo {author} {\bibfnamefont {F.~X.}\ \bibnamefont {Girod}}, \bibinfo {author} {\bibfnamefont {C.}~\bibnamefont {Lorc\'e}}, \bibinfo {author} {\bibfnamefont {P.}~\bibnamefont {Schweitzer}},\ and\ \bibinfo {author} {\bibfnamefont {P.~E.}\ \bibnamefont {Shanahan}},\ }\bibfield  {title} {\bibinfo {title} {{Colloquium: Gravitational Form Factors of the Proton}},\ }\href@noop {} {\  (\bibinfo {year} {2023})},\ \Eprint {https://arxiv.org/abs/2303.08347} {arXiv:2303.08347 [hep-ph]} \BibitemShut {NoStop}%
\bibitem [{\citenamefont {Chen}\ \emph {et~al.}(2006)\citenamefont {Chen} \emph {et~al.}}]{Chen:2005mg}%
  \BibitemOpen
  \bibfield  {author} {\bibinfo {author} {\bibfnamefont {Y.}~\bibnamefont {Chen}} \emph {et~al.},\ }\bibfield  {title} {\bibinfo {title} {{Glueball spectrum and matrix elements on anisotropic lattices}},\ }\href {https://doi.org/10.1103/PhysRevD.73.014516} {\bibfield  {journal} {\bibinfo  {journal} {Phys. Rev. D}\ }\textbf {\bibinfo {volume} {73}},\ \bibinfo {pages} {014516} (\bibinfo {year} {2006})},\ \Eprint {https://arxiv.org/abs/hep-lat/0510074} {arXiv:hep-lat/0510074} \BibitemShut {NoStop}%
\bibitem [{\citenamefont {Abbott}\ \emph {et~al.}(2025{\natexlab{a}})\citenamefont {Abbott}, \citenamefont {Hackett}, \citenamefont {Pefkou}, \citenamefont {Romero-L{\'o}pez},\ and\ \citenamefont {Shanahan}}]{Abbott:2024bre}%
  \BibitemOpen
  \bibfield  {author} {\bibinfo {author} {\bibfnamefont {R.}~\bibnamefont {Abbott}}, \bibinfo {author} {\bibfnamefont {D.~C.}\ \bibnamefont {Hackett}}, \bibinfo {author} {\bibfnamefont {D.~A.}\ \bibnamefont {Pefkou}}, \bibinfo {author} {\bibfnamefont {F.}~\bibnamefont {Romero-L{\'o}pez}},\ and\ \bibinfo {author} {\bibfnamefont {P.}~\bibnamefont {Shanahan}},\ }\bibfield  {title} {\bibinfo {title} {{Gravitational form factors of glueballs in Yang-Mills theory}},\ }\href {https://doi.org/10.22323/1.466.0459} {\bibfield  {journal} {\bibinfo  {journal} {PoS}\ }\textbf {\bibinfo {volume} {LATTICE2024}},\ \bibinfo {pages} {459} (\bibinfo {year} {2025}{\natexlab{a}})},\ \Eprint {https://arxiv.org/abs/2410.02706} {arXiv:2410.02706 [hep-lat]} \BibitemShut {NoStop}%
\bibitem [{\citenamefont {Pagels}(1966)}]{Pagels:1966zza}%
  \BibitemOpen
  \bibfield  {author} {\bibinfo {author} {\bibfnamefont {H.}~\bibnamefont {Pagels}},\ }\bibfield  {title} {\bibinfo {title} {{Energy-Momentum Structure Form Factors of Particles}},\ }\href {https://doi.org/10.1103/PhysRev.144.1250} {\bibfield  {journal} {\bibinfo  {journal} {Phys. Rev.}\ }\textbf {\bibinfo {volume} {144}},\ \bibinfo {pages} {1250} (\bibinfo {year} {1966})}\BibitemShut {NoStop}%
\bibitem [{\citenamefont {Hudson}\ and\ \citenamefont {Schweitzer}(2017)}]{Hudson:2017xug}%
  \BibitemOpen
  \bibfield  {author} {\bibinfo {author} {\bibfnamefont {J.}~\bibnamefont {Hudson}}\ and\ \bibinfo {author} {\bibfnamefont {P.}~\bibnamefont {Schweitzer}},\ }\bibfield  {title} {\bibinfo {title} {{D term and the structure of pointlike and composed spin-0 particles}},\ }\href {https://doi.org/10.1103/PhysRevD.96.114013} {\bibfield  {journal} {\bibinfo  {journal} {Phys. Rev. D}\ }\textbf {\bibinfo {volume} {96}},\ \bibinfo {pages} {114013} (\bibinfo {year} {2017})},\ \Eprint {https://arxiv.org/abs/1712.05316} {arXiv:1712.05316 [hep-ph]} \BibitemShut {NoStop}%
\bibitem [{\citenamefont {Cotogno}\ \emph {et~al.}(2020)\citenamefont {Cotogno}, \citenamefont {Lorc{\'e}}, \citenamefont {Lowdon},\ and\ \citenamefont {Morales}}]{Cotogno:2019vjb}%
  \BibitemOpen
  \bibfield  {author} {\bibinfo {author} {\bibfnamefont {S.}~\bibnamefont {Cotogno}}, \bibinfo {author} {\bibfnamefont {C.}~\bibnamefont {Lorc{\'e}}}, \bibinfo {author} {\bibfnamefont {P.}~\bibnamefont {Lowdon}},\ and\ \bibinfo {author} {\bibfnamefont {M.}~\bibnamefont {Morales}},\ }\bibfield  {title} {\bibinfo {title} {{Covariant multipole expansion of local currents for massive states of any spin}},\ }\href {https://doi.org/10.1103/PhysRevD.101.056016} {\bibfield  {journal} {\bibinfo  {journal} {Phys. Rev. D}\ }\textbf {\bibinfo {volume} {101}},\ \bibinfo {pages} {056016} (\bibinfo {year} {2020})},\ \Eprint {https://arxiv.org/abs/1912.08749} {arXiv:1912.08749 [hep-ph]} \BibitemShut {NoStop}%
\bibitem [{\citenamefont {Necco}\ and\ \citenamefont {Sommer}(2002)}]{Necco:2001xg}%
  \BibitemOpen
  \bibfield  {author} {\bibinfo {author} {\bibfnamefont {S.}~\bibnamefont {Necco}}\ and\ \bibinfo {author} {\bibfnamefont {R.}~\bibnamefont {Sommer}},\ }\bibfield  {title} {\bibinfo {title} {{The N(f) = 0 heavy quark potential from short to intermediate distances}},\ }\href {https://doi.org/10.1016/S0550-3213(01)00582-X} {\bibfield  {journal} {\bibinfo  {journal} {Nucl. Phys. B}\ }\textbf {\bibinfo {volume} {622}},\ \bibinfo {pages} {328} (\bibinfo {year} {2002})},\ \Eprint {https://arxiv.org/abs/hep-lat/0108008} {arXiv:hep-lat/0108008} \BibitemShut {NoStop}%
\bibitem [{\citenamefont {Durr}\ \emph {et~al.}(2007)\citenamefont {Durr}, \citenamefont {Fodor}, \citenamefont {Hoelbling},\ and\ \citenamefont {Kurth}}]{Durr:2006ky}%
  \BibitemOpen
  \bibfield  {author} {\bibinfo {author} {\bibfnamefont {S.}~\bibnamefont {Durr}}, \bibinfo {author} {\bibfnamefont {Z.}~\bibnamefont {Fodor}}, \bibinfo {author} {\bibfnamefont {C.}~\bibnamefont {Hoelbling}},\ and\ \bibinfo {author} {\bibfnamefont {T.}~\bibnamefont {Kurth}},\ }\bibfield  {title} {\bibinfo {title} {{Precision study of the SU(3) topological susceptibility in the continuum}},\ }\href {https://doi.org/10.1088/1126-6708/2007/04/055} {\bibfield  {journal} {\bibinfo  {journal} {JHEP}\ }\textbf {\bibinfo {volume} {04}},\ \bibinfo {pages} {055}},\ \Eprint {https://arxiv.org/abs/hep-lat/0612021} {arXiv:hep-lat/0612021} \BibitemShut {NoStop}%
\bibitem [{\citenamefont {Creutz}(1980)}]{Creutz:1980zw}%
  \BibitemOpen
  \bibfield  {author} {\bibinfo {author} {\bibfnamefont {M.}~\bibnamefont {Creutz}},\ }\bibfield  {title} {\bibinfo {title} {{Monte Carlo Study of Quantized SU(2) Gauge Theory}},\ }\href {https://doi.org/10.1103/PhysRevD.21.2308} {\bibfield  {journal} {\bibinfo  {journal} {Phys. Rev. D}\ }\textbf {\bibinfo {volume} {21}},\ \bibinfo {pages} {2308} (\bibinfo {year} {1980})}\BibitemShut {NoStop}%
\bibitem [{\citenamefont {Cabibbo}\ and\ \citenamefont {Marinari}(1982)}]{Cabibbo:1982zn}%
  \BibitemOpen
  \bibfield  {author} {\bibinfo {author} {\bibfnamefont {N.}~\bibnamefont {Cabibbo}}\ and\ \bibinfo {author} {\bibfnamefont {E.}~\bibnamefont {Marinari}},\ }\bibfield  {title} {\bibinfo {title} {{A New Method for Updating SU(N) Matrices in Computer Simulations of Gauge Theories}},\ }\href {https://doi.org/10.1016/0370-2693(82)90696-7} {\bibfield  {journal} {\bibinfo  {journal} {Phys. Lett. B}\ }\textbf {\bibinfo {volume} {119}},\ \bibinfo {pages} {387} (\bibinfo {year} {1982})}\BibitemShut {NoStop}%
\bibitem [{\citenamefont {Kennedy}\ and\ \citenamefont {Pendleton}(1985)}]{Kennedy:1985nu}%
  \BibitemOpen
  \bibfield  {author} {\bibinfo {author} {\bibfnamefont {A.~D.}\ \bibnamefont {Kennedy}}\ and\ \bibinfo {author} {\bibfnamefont {B.~J.}\ \bibnamefont {Pendleton}},\ }\bibfield  {title} {\bibinfo {title} {{Improved Heat Bath Method for Monte Carlo Calculations in Lattice Gauge Theories}},\ }\href {https://doi.org/10.1016/0370-2693(85)91632-6} {\bibfield  {journal} {\bibinfo  {journal} {Phys. Lett. B}\ }\textbf {\bibinfo {volume} {156}},\ \bibinfo {pages} {393} (\bibinfo {year} {1985})}\BibitemShut {NoStop}%
\bibitem [{\citenamefont {Brown}\ and\ \citenamefont {Woch}(1987)}]{Brown:1987rra}%
  \BibitemOpen
  \bibfield  {author} {\bibinfo {author} {\bibfnamefont {F.~R.}\ \bibnamefont {Brown}}\ and\ \bibinfo {author} {\bibfnamefont {T.~J.}\ \bibnamefont {Woch}},\ }\bibfield  {title} {\bibinfo {title} {{Overrelaxed Heat Bath and Metropolis Algorithms for Accelerating Pure Gauge Monte Carlo Calculations}},\ }\href {https://doi.org/10.1103/PhysRevLett.58.2394} {\bibfield  {journal} {\bibinfo  {journal} {Phys. Rev. Lett.}\ }\textbf {\bibinfo {volume} {58}},\ \bibinfo {pages} {2394} (\bibinfo {year} {1987})}\BibitemShut {NoStop}%
\bibitem [{\citenamefont {Adler}(1988)}]{Adler:1987ce}%
  \BibitemOpen
  \bibfield  {author} {\bibinfo {author} {\bibfnamefont {S.~L.}\ \bibnamefont {Adler}},\ }\bibfield  {title} {\bibinfo {title} {{Overrelaxation Algorithms for Lattice Field Theories}},\ }\href {https://doi.org/10.1103/PhysRevD.37.458} {\bibfield  {journal} {\bibinfo  {journal} {Phys. Rev. D}\ }\textbf {\bibinfo {volume} {37}},\ \bibinfo {pages} {458} (\bibinfo {year} {1988})}\BibitemShut {NoStop}%
\bibitem [{\citenamefont {Morningstar}\ and\ \citenamefont {Peardon}(2004)}]{Morningstar:2003gk}%
  \BibitemOpen
  \bibfield  {author} {\bibinfo {author} {\bibfnamefont {C.}~\bibnamefont {Morningstar}}\ and\ \bibinfo {author} {\bibfnamefont {M.~J.}\ \bibnamefont {Peardon}},\ }\bibfield  {title} {\bibinfo {title} {{Analytic smearing of SU(3) link variables in lattice QCD}},\ }\href {https://doi.org/10.1103/PhysRevD.69.054501} {\bibfield  {journal} {\bibinfo  {journal} {Phys. Rev. D}\ }\textbf {\bibinfo {volume} {69}},\ \bibinfo {pages} {054501} (\bibinfo {year} {2004})},\ \Eprint {https://arxiv.org/abs/hep-lat/0311018} {arXiv:hep-lat/0311018} \BibitemShut {NoStop}%
\bibitem [{\citenamefont {Michael}(1989)}]{Michael:1989vh}%
  \BibitemOpen
  \bibfield  {author} {\bibinfo {author} {\bibfnamefont {C.}~\bibnamefont {Michael}},\ }\bibfield  {title} {\bibinfo {title} {{Torelons and Unusual Ground States}},\ }\href {https://doi.org/10.1016/0370-2693(89)91695-X} {\bibfield  {journal} {\bibinfo  {journal} {Phys. Lett. B}\ }\textbf {\bibinfo {volume} {232}},\ \bibinfo {pages} {247} (\bibinfo {year} {1989})}\BibitemShut {NoStop}%
\bibitem [{\citenamefont {Fox}\ \emph {et~al.}(1982)\citenamefont {Fox}, \citenamefont {Gupta}, \citenamefont {Martin},\ and\ \citenamefont {Otto}}]{Fox:1981xz}%
  \BibitemOpen
  \bibfield  {author} {\bibinfo {author} {\bibfnamefont {G.}~\bibnamefont {Fox}}, \bibinfo {author} {\bibfnamefont {R.}~\bibnamefont {Gupta}}, \bibinfo {author} {\bibfnamefont {O.}~\bibnamefont {Martin}},\ and\ \bibinfo {author} {\bibfnamefont {S.}~\bibnamefont {Otto}},\ }\bibfield  {title} {\bibinfo {title} {{Monte Carlo Estimates of the Mass Gap of the O(2) and O(3) Spin Models in (1+1)-dimensions}},\ }\href {https://doi.org/10.1016/0550-3213(82)90384-4} {\bibfield  {journal} {\bibinfo  {journal} {Nucl. Phys. B}\ }\textbf {\bibinfo {volume} {205}},\ \bibinfo {pages} {188} (\bibinfo {year} {1982})}\BibitemShut {NoStop}%
\bibitem [{\citenamefont {Michael}\ and\ \citenamefont {Teasdale}(1983)}]{Michael:1982gb}%
  \BibitemOpen
  \bibfield  {author} {\bibinfo {author} {\bibfnamefont {C.}~\bibnamefont {Michael}}\ and\ \bibinfo {author} {\bibfnamefont {I.}~\bibnamefont {Teasdale}},\ }\bibfield  {title} {\bibinfo {title} {{Extracting Glueball Masses From Lattice {QCD}}},\ }\href {https://doi.org/10.1016/0550-3213(83)90674-0} {\bibfield  {journal} {\bibinfo  {journal} {Nucl. Phys. B}\ }\textbf {\bibinfo {volume} {215}},\ \bibinfo {pages} {433} (\bibinfo {year} {1983})}\BibitemShut {NoStop}%
\bibitem [{\citenamefont {Luscher}\ and\ \citenamefont {Wolff}(1990)}]{Luscher:1990ck}%
  \BibitemOpen
  \bibfield  {author} {\bibinfo {author} {\bibfnamefont {M.}~\bibnamefont {Luscher}}\ and\ \bibinfo {author} {\bibfnamefont {U.}~\bibnamefont {Wolff}},\ }\bibfield  {title} {\bibinfo {title} {{How to Calculate the Elastic Scattering Matrix in Two-dimensional Quantum Field Theories by Numerical Simulation}},\ }\href {https://doi.org/10.1016/0550-3213(90)90540-T} {\bibfield  {journal} {\bibinfo  {journal} {Nucl. Phys. B}\ }\textbf {\bibinfo {volume} {339}},\ \bibinfo {pages} {222} (\bibinfo {year} {1990})}\BibitemShut {NoStop}%
\bibitem [{\citenamefont {Blossier}\ \emph {et~al.}(2009)\citenamefont {Blossier}, \citenamefont {Della~Morte}, \citenamefont {von Hippel}, \citenamefont {Mendes},\ and\ \citenamefont {Sommer}}]{Blossier:2009kd}%
  \BibitemOpen
  \bibfield  {author} {\bibinfo {author} {\bibfnamefont {B.}~\bibnamefont {Blossier}}, \bibinfo {author} {\bibfnamefont {M.}~\bibnamefont {Della~Morte}}, \bibinfo {author} {\bibfnamefont {G.}~\bibnamefont {von Hippel}}, \bibinfo {author} {\bibfnamefont {T.}~\bibnamefont {Mendes}},\ and\ \bibinfo {author} {\bibfnamefont {R.}~\bibnamefont {Sommer}},\ }\bibfield  {title} {\bibinfo {title} {{On the generalized eigenvalue method for energies and matrix elements in lattice field theory}},\ }\href {https://doi.org/10.1088/1126-6708/2009/04/094} {\bibfield  {journal} {\bibinfo  {journal} {JHEP}\ }\textbf {\bibinfo {volume} {04}},\ \bibinfo {pages} {094}},\ \Eprint {https://arxiv.org/abs/0902.1265} {arXiv:0902.1265 [hep-lat]} \BibitemShut {NoStop}%
\bibitem [{\citenamefont {Fleming}(2023)}]{Fleming:2023zml}%
  \BibitemOpen
  \bibfield  {author} {\bibinfo {author} {\bibfnamefont {G.~T.}\ \bibnamefont {Fleming}},\ }\bibfield  {title} {\bibinfo {title} {{Beyond Generalized Eigenvalues in Lattice Quantum Field Theory}},\ }in\ \href@noop {} {\emph {\bibinfo {booktitle} {{40th International Symposium on Lattice Field Theory}}}}\ (\bibinfo {year} {2023})\ \Eprint {https://arxiv.org/abs/2309.05111} {arXiv:2309.05111 [hep-lat]} \BibitemShut {NoStop}%
\bibitem [{\citenamefont {Fischer}\ \emph {et~al.}(2020)\citenamefont {Fischer}, \citenamefont {Kostrzewa}, \citenamefont {Ostmeyer}, \citenamefont {Ottnad}, \citenamefont {Ueding},\ and\ \citenamefont {Urbach}}]{Fischer:2020bgv}%
  \BibitemOpen
  \bibfield  {author} {\bibinfo {author} {\bibfnamefont {M.}~\bibnamefont {Fischer}}, \bibinfo {author} {\bibfnamefont {B.}~\bibnamefont {Kostrzewa}}, \bibinfo {author} {\bibfnamefont {J.}~\bibnamefont {Ostmeyer}}, \bibinfo {author} {\bibfnamefont {K.}~\bibnamefont {Ottnad}}, \bibinfo {author} {\bibfnamefont {M.}~\bibnamefont {Ueding}},\ and\ \bibinfo {author} {\bibfnamefont {C.}~\bibnamefont {Urbach}},\ }\bibfield  {title} {\bibinfo {title} {{On the generalised eigenvalue method and its relation to Prony and generalised pencil of function methods}},\ }\href {https://doi.org/10.1140/epja/s10050-020-00205-w} {\bibfield  {journal} {\bibinfo  {journal} {Eur. Phys. J. A}\ }\textbf {\bibinfo {volume} {56}},\ \bibinfo {pages} {206} (\bibinfo {year} {2020})},\ \Eprint {https://arxiv.org/abs/2004.10472} {arXiv:2004.10472 [hep-lat]} \BibitemShut {NoStop}%
\bibitem [{\citenamefont {Sterman}\ and\ \citenamefont {Vogelsang}(2000)}]{Sterman:2000pu}%
  \BibitemOpen
  \bibfield  {author} {\bibinfo {author} {\bibfnamefont {G.~F.}\ \bibnamefont {Sterman}}\ and\ \bibinfo {author} {\bibfnamefont {W.}~\bibnamefont {Vogelsang}},\ }\bibfield  {title} {\bibinfo {title} {{Soft gluon resummation and PDF theory uncertainties}},\ }in\ \href@noop {} {\emph {\bibinfo {booktitle} {{Physics at Run II: QCD and Weak Boson Physics Workshop: Final General Meeting}}}}\ (\bibinfo {year} {2000})\ \Eprint {https://arxiv.org/abs/hep-ph/0002132} {arXiv:hep-ph/0002132} \BibitemShut {NoStop}%
\bibitem [{\citenamefont {Pefkou}\ \emph {et~al.}(2022)\citenamefont {Pefkou}, \citenamefont {Hackett},\ and\ \citenamefont {Shanahan}}]{Pefkou:2021fni}%
  \BibitemOpen
  \bibfield  {author} {\bibinfo {author} {\bibfnamefont {D.~A.}\ \bibnamefont {Pefkou}}, \bibinfo {author} {\bibfnamefont {D.~C.}\ \bibnamefont {Hackett}},\ and\ \bibinfo {author} {\bibfnamefont {P.~E.}\ \bibnamefont {Shanahan}},\ }\bibfield  {title} {\bibinfo {title} {{Gluon gravitational structure of hadrons of different spin}},\ }\href {https://doi.org/10.1103/PhysRevD.105.054509} {\bibfield  {journal} {\bibinfo  {journal} {Phys. Rev. D}\ }\textbf {\bibinfo {volume} {105}},\ \bibinfo {pages} {054509} (\bibinfo {year} {2022})},\ \Eprint {https://arxiv.org/abs/2107.10368} {arXiv:2107.10368 [hep-lat]} \BibitemShut {NoStop}%
\bibitem [{\citenamefont {Hackett}\ \emph {et~al.}(2024)\citenamefont {Hackett}, \citenamefont {Pefkou},\ and\ \citenamefont {Shanahan}}]{Hackett:2023rif}%
  \BibitemOpen
  \bibfield  {author} {\bibinfo {author} {\bibfnamefont {D.~C.}\ \bibnamefont {Hackett}}, \bibinfo {author} {\bibfnamefont {D.~A.}\ \bibnamefont {Pefkou}},\ and\ \bibinfo {author} {\bibfnamefont {P.~E.}\ \bibnamefont {Shanahan}},\ }\bibfield  {title} {\bibinfo {title} {{Gravitational Form Factors of the Proton from Lattice QCD}},\ }\href {https://doi.org/10.1103/PhysRevLett.132.251904} {\bibfield  {journal} {\bibinfo  {journal} {Phys. Rev. Lett.}\ }\textbf {\bibinfo {volume} {132}},\ \bibinfo {pages} {251904} (\bibinfo {year} {2024})},\ \Eprint {https://arxiv.org/abs/2310.08484} {arXiv:2310.08484 [hep-lat]} \BibitemShut {NoStop}%
\bibitem [{\citenamefont {Hackett}\ \emph {et~al.}(2023)\citenamefont {Hackett}, \citenamefont {Oare}, \citenamefont {Pefkou},\ and\ \citenamefont {Shanahan}}]{Hackett:2023nkr}%
  \BibitemOpen
  \bibfield  {author} {\bibinfo {author} {\bibfnamefont {D.~C.}\ \bibnamefont {Hackett}}, \bibinfo {author} {\bibfnamefont {P.~R.}\ \bibnamefont {Oare}}, \bibinfo {author} {\bibfnamefont {D.~A.}\ \bibnamefont {Pefkou}},\ and\ \bibinfo {author} {\bibfnamefont {P.~E.}\ \bibnamefont {Shanahan}},\ }\bibfield  {title} {\bibinfo {title} {{Gravitational form factors of the pion from lattice QCD}},\ }\href {https://doi.org/10.1103/PhysRevD.108.114504} {\bibfield  {journal} {\bibinfo  {journal} {Phys. Rev. D}\ }\textbf {\bibinfo {volume} {108}},\ \bibinfo {pages} {114504} (\bibinfo {year} {2023})},\ \Eprint {https://arxiv.org/abs/2307.11707} {arXiv:2307.11707 [hep-lat]} \BibitemShut {NoStop}%
\bibitem [{\citenamefont {Gockeler}\ \emph {et~al.}(1996)\citenamefont {Gockeler}, \citenamefont {Horsley}, \citenamefont {Ilgenfritz}, \citenamefont {Perlt}, \citenamefont {Rakow}, \citenamefont {Schierholz},\ and\ \citenamefont {Schiller}}]{Gockeler:1996mu}%
  \BibitemOpen
  \bibfield  {author} {\bibinfo {author} {\bibfnamefont {M.}~\bibnamefont {Gockeler}}, \bibinfo {author} {\bibfnamefont {R.}~\bibnamefont {Horsley}}, \bibinfo {author} {\bibfnamefont {E.-M.}\ \bibnamefont {Ilgenfritz}}, \bibinfo {author} {\bibfnamefont {H.}~\bibnamefont {Perlt}}, \bibinfo {author} {\bibfnamefont {P.~E.~L.}\ \bibnamefont {Rakow}}, \bibinfo {author} {\bibfnamefont {G.}~\bibnamefont {Schierholz}},\ and\ \bibinfo {author} {\bibfnamefont {A.}~\bibnamefont {Schiller}},\ }\bibfield  {title} {\bibinfo {title} {{Lattice operators for moments of the structure functions and their transformation under the hypercubic group}},\ }\href {https://doi.org/10.1103/PhysRevD.54.5705} {\bibfield  {journal} {\bibinfo  {journal} {Phys. Rev. D}\ }\textbf {\bibinfo {volume} {54}},\ \bibinfo {pages} {5705} (\bibinfo {year} {1996})},\ \Eprint {https://arxiv.org/abs/hep-lat/9602029} {arXiv:hep-lat/9602029} \BibitemShut {NoStop}%
\bibitem [{\citenamefont {Wagman}(2025)}]{Wagman:2024rid}%
  \BibitemOpen
  \bibfield  {author} {\bibinfo {author} {\bibfnamefont {M.~L.}\ \bibnamefont {Wagman}},\ }\bibfield  {title} {\bibinfo {title} {{Lanczos Algorithm, the Transfer Matrix, and the Signal-to-Noise Problem}},\ }\href {https://doi.org/10.1103/pcvc-734h} {\bibfield  {journal} {\bibinfo  {journal} {Phys. Rev. Lett.}\ }\textbf {\bibinfo {volume} {134}},\ \bibinfo {pages} {241901} (\bibinfo {year} {2025})},\ \Eprint {https://arxiv.org/abs/2406.20009} {arXiv:2406.20009 [hep-lat]} \BibitemShut {NoStop}%
\bibitem [{\citenamefont {Hackett}\ and\ \citenamefont {Wagman}(2024)}]{Hackett:2024xnx}%
  \BibitemOpen
  \bibfield  {author} {\bibinfo {author} {\bibfnamefont {D.~C.}\ \bibnamefont {Hackett}}\ and\ \bibinfo {author} {\bibfnamefont {M.~L.}\ \bibnamefont {Wagman}},\ }\bibfield  {title} {\bibinfo {title} {{Lanczos for lattice QCD matrix elements}},\ }\href@noop {} {\  (\bibinfo {year} {2024})},\ \Eprint {https://arxiv.org/abs/2407.21777} {arXiv:2407.21777 [hep-lat]} \BibitemShut {NoStop}%
\bibitem [{\citenamefont {Hackett}\ and\ \citenamefont {Wagman}(2025)}]{Hackett:2024nbe}%
  \BibitemOpen
  \bibfield  {author} {\bibinfo {author} {\bibfnamefont {D.~C.}\ \bibnamefont {Hackett}}\ and\ \bibinfo {author} {\bibfnamefont {M.~L.}\ \bibnamefont {Wagman}},\ }\bibfield  {title} {\bibinfo {title} {{Block Lanczos algorithm for lattice QCD spectroscopy and matrix elements}},\ }\href {https://doi.org/10.1103/fp74-q35q} {\bibfield  {journal} {\bibinfo  {journal} {Phys. Rev. D}\ }\textbf {\bibinfo {volume} {112}},\ \bibinfo {pages} {014514} (\bibinfo {year} {2025})},\ \Eprint {https://arxiv.org/abs/2412.04444} {arXiv:2412.04444 [hep-lat]} \BibitemShut {NoStop}%
\bibitem [{\citenamefont {Abbott}\ \emph {et~al.}(2025{\natexlab{b}})\citenamefont {Abbott}, \citenamefont {Hackett}, \citenamefont {Fleming}, \citenamefont {Pefkou},\ and\ \citenamefont {Wagman}}]{Abbott:2025yhm}%
  \BibitemOpen
  \bibfield  {author} {\bibinfo {author} {\bibfnamefont {R.}~\bibnamefont {Abbott}}, \bibinfo {author} {\bibfnamefont {D.~C.}\ \bibnamefont {Hackett}}, \bibinfo {author} {\bibfnamefont {G.~T.}\ \bibnamefont {Fleming}}, \bibinfo {author} {\bibfnamefont {D.~A.}\ \bibnamefont {Pefkou}},\ and\ \bibinfo {author} {\bibfnamefont {M.~L.}\ \bibnamefont {Wagman}},\ }\bibfield  {title} {\bibinfo {title} {{Filtered Rayleigh-Ritz is all you need}},\ }\href@noop {} {\  (\bibinfo {year} {2025}{\natexlab{b}})},\ \Eprint {https://arxiv.org/abs/2503.17357} {arXiv:2503.17357 [hep-lat]} \BibitemShut {NoStop}%
\bibitem [{\citenamefont {Br{\"o}mmel}(2007)}]{Brommel:2007zz}%
  \BibitemOpen
  \bibfield  {author} {\bibinfo {author} {\bibfnamefont {D.}~\bibnamefont {Br{\"o}mmel}},\ }\emph {\bibinfo {title} {{Pion Structure from the Lattice}}},\ \href {https://doi.org/10.3204/DESY-THESIS-2007-023} {Ph.D. thesis},\ \bibinfo  {school} {Regensburg U.} (\bibinfo {year} {2007})\BibitemShut {NoStop}%
\bibitem [{\citenamefont {Detmold}\ \emph {et~al.}(2017)\citenamefont {Detmold}, \citenamefont {Pefkou},\ and\ \citenamefont {Shanahan}}]{Detmold:2017oqb}%
  \BibitemOpen
  \bibfield  {author} {\bibinfo {author} {\bibfnamefont {W.}~\bibnamefont {Detmold}}, \bibinfo {author} {\bibfnamefont {D.}~\bibnamefont {Pefkou}},\ and\ \bibinfo {author} {\bibfnamefont {P.~E.}\ \bibnamefont {Shanahan}},\ }\bibfield  {title} {\bibinfo {title} {{Off-forward gluonic structure of vector mesons}},\ }\href {https://doi.org/10.1103/PhysRevD.95.114515} {\bibfield  {journal} {\bibinfo  {journal} {Phys. Rev. D}\ }\textbf {\bibinfo {volume} {95}},\ \bibinfo {pages} {114515} (\bibinfo {year} {2017})},\ \Eprint {https://arxiv.org/abs/1703.08220} {arXiv:1703.08220 [hep-lat]} \BibitemShut {NoStop}%
\bibitem [{\citenamefont {Shanahan}\ and\ \citenamefont {Detmold}(2019{\natexlab{a}})}]{Shanahan:2018pib}%
  \BibitemOpen
  \bibfield  {author} {\bibinfo {author} {\bibfnamefont {P.}~\bibnamefont {Shanahan}}\ and\ \bibinfo {author} {\bibfnamefont {W.}~\bibnamefont {Detmold}},\ }\bibfield  {title} {\bibinfo {title} {{Gluon gravitational form factors of the nucleon and the pion from lattice QCD}},\ }\href {https://doi.org/10.1103/PhysRevD.99.014511} {\bibfield  {journal} {\bibinfo  {journal} {Phys. Rev. D}\ }\textbf {\bibinfo {volume} {99}},\ \bibinfo {pages} {014511} (\bibinfo {year} {2019}{\natexlab{a}})},\ \Eprint {https://arxiv.org/abs/1810.04626} {arXiv:1810.04626 [hep-lat]} \BibitemShut {NoStop}%
\bibitem [{\citenamefont {Shanahan}\ and\ \citenamefont {Detmold}(2019{\natexlab{b}})}]{Shanahan:2018nnv}%
  \BibitemOpen
  \bibfield  {author} {\bibinfo {author} {\bibfnamefont {P.}~\bibnamefont {Shanahan}}\ and\ \bibinfo {author} {\bibfnamefont {W.}~\bibnamefont {Detmold}},\ }\bibfield  {title} {\bibinfo {title} {{Pressure Distribution and Shear Forces inside the Proton}},\ }\href {https://doi.org/10.1103/PhysRevLett.122.072003} {\bibfield  {journal} {\bibinfo  {journal} {Phys. Rev. Lett.}\ }\textbf {\bibinfo {volume} {122}},\ \bibinfo {pages} {072003} (\bibinfo {year} {2019}{\natexlab{b}})},\ \Eprint {https://arxiv.org/abs/1810.07589} {arXiv:1810.07589 [nucl-th]} \BibitemShut {NoStop}%
\bibitem [{\citenamefont {Steinberg}(2019)}]{kmeans1d}%
  \BibitemOpen
  \bibfield  {author} {\bibinfo {author} {\bibfnamefont {D.}~\bibnamefont {Steinberg}},\ }\href@noop {} {\bibinfo {title} {kmeans1d}},\ \bibinfo {howpublished} {\url{https://github.com/dstein64/kmeans1d}} (\bibinfo {year} {2019})\BibitemShut {NoStop}%
\bibitem [{\citenamefont {Freese}\ and\ \citenamefont {Clo{\"e}t}(2019)}]{Freese:2019bhb}%
  \BibitemOpen
  \bibfield  {author} {\bibinfo {author} {\bibfnamefont {A.}~\bibnamefont {Freese}}\ and\ \bibinfo {author} {\bibfnamefont {I.~C.}\ \bibnamefont {Clo{\"e}t}},\ }\bibfield  {title} {\bibinfo {title} {{Gravitational form factors of light mesons}},\ }\href {https://doi.org/10.1103/PhysRevC.100.015201} {\bibfield  {journal} {\bibinfo  {journal} {Phys. Rev. C}\ }\textbf {\bibinfo {volume} {100}},\ \bibinfo {pages} {015201} (\bibinfo {year} {2019})},\ \bibinfo {note} {[Erratum: Phys.Rev.C 105, 059901 (2022)]},\ \Eprint {https://arxiv.org/abs/1903.09222} {arXiv:1903.09222 [nucl-th]} \BibitemShut {NoStop}%
\bibitem [{\citenamefont {Yennie}\ \emph {et~al.}(1957)\citenamefont {Yennie}, \citenamefont {L{\'e}vy},\ and\ \citenamefont {Ravenhall}}]{Yennie:1957skg}%
  \BibitemOpen
  \bibfield  {author} {\bibinfo {author} {\bibfnamefont {D.~R.}\ \bibnamefont {Yennie}}, \bibinfo {author} {\bibfnamefont {M.~M.}\ \bibnamefont {L{\'e}vy}},\ and\ \bibinfo {author} {\bibfnamefont {D.~G.}\ \bibnamefont {Ravenhall}},\ }\bibfield  {title} {\bibinfo {title} {{Electromagnetic Structure of Nucleons}},\ }\href {https://doi.org/10.1103/RevModPhys.29.144} {\bibfield  {journal} {\bibinfo  {journal} {Rev. Mod. Phys.}\ }\textbf {\bibinfo {volume} {29}},\ \bibinfo {pages} {144} (\bibinfo {year} {1957})}\BibitemShut {NoStop}%
\bibitem [{\citenamefont {Burkardt}(2003)}]{Burkardt:2002hr}%
  \BibitemOpen
  \bibfield  {author} {\bibinfo {author} {\bibfnamefont {M.}~\bibnamefont {Burkardt}},\ }\bibfield  {title} {\bibinfo {title} {{Impact parameter space interpretation for generalized parton distributions}},\ }\href {https://doi.org/10.1142/S0217751X03012370} {\bibfield  {journal} {\bibinfo  {journal} {Int. J. Mod. Phys. A}\ }\textbf {\bibinfo {volume} {18}},\ \bibinfo {pages} {173} (\bibinfo {year} {2003})},\ \Eprint {https://arxiv.org/abs/hep-ph/0207047} {arXiv:hep-ph/0207047} \BibitemShut {NoStop}%
\bibitem [{\citenamefont {Miller}(2019)}]{Miller:2018ybm}%
  \BibitemOpen
  \bibfield  {author} {\bibinfo {author} {\bibfnamefont {G.~A.}\ \bibnamefont {Miller}},\ }\bibfield  {title} {\bibinfo {title} {{Defining the proton radius: A unified treatment}},\ }\href {https://doi.org/10.1103/PhysRevC.99.035202} {\bibfield  {journal} {\bibinfo  {journal} {Phys. Rev. C}\ }\textbf {\bibinfo {volume} {99}},\ \bibinfo {pages} {035202} (\bibinfo {year} {2019})},\ \Eprint {https://arxiv.org/abs/1812.02714} {arXiv:1812.02714 [nucl-th]} \BibitemShut {NoStop}%
\bibitem [{\citenamefont {Jaffe}(2021)}]{Jaffe:2020ebz}%
  \BibitemOpen
  \bibfield  {author} {\bibinfo {author} {\bibfnamefont {R.~L.}\ \bibnamefont {Jaffe}},\ }\bibfield  {title} {\bibinfo {title} {{Ambiguities in the definition of local spatial densities in light hadrons}},\ }\href {https://doi.org/10.1103/PhysRevD.103.016017} {\bibfield  {journal} {\bibinfo  {journal} {Phys. Rev. D}\ }\textbf {\bibinfo {volume} {103}},\ \bibinfo {pages} {016017} (\bibinfo {year} {2021})},\ \Eprint {https://arxiv.org/abs/2010.15887} {arXiv:2010.15887 [hep-ph]} \BibitemShut {NoStop}%
\bibitem [{\citenamefont {Epelbaum}\ \emph {et~al.}(2022)\citenamefont {Epelbaum}, \citenamefont {Gegelia}, \citenamefont {Lange}, \citenamefont {Mei{\ss}ner},\ and\ \citenamefont {Polyakov}}]{Epelbaum:2022fjc}%
  \BibitemOpen
  \bibfield  {author} {\bibinfo {author} {\bibfnamefont {E.}~\bibnamefont {Epelbaum}}, \bibinfo {author} {\bibfnamefont {J.}~\bibnamefont {Gegelia}}, \bibinfo {author} {\bibfnamefont {N.}~\bibnamefont {Lange}}, \bibinfo {author} {\bibfnamefont {U.~G.}\ \bibnamefont {Mei{\ss}ner}},\ and\ \bibinfo {author} {\bibfnamefont {M.~V.}\ \bibnamefont {Polyakov}},\ }\bibfield  {title} {\bibinfo {title} {{Definition of Local Spatial Densities in Hadrons}},\ }\href {https://doi.org/10.1103/PhysRevLett.129.012001} {\bibfield  {journal} {\bibinfo  {journal} {Phys. Rev. Lett.}\ }\textbf {\bibinfo {volume} {129}},\ \bibinfo {pages} {012001} (\bibinfo {year} {2022})},\ \Eprint {https://arxiv.org/abs/2201.02565} {arXiv:2201.02565 [hep-ph]} \BibitemShut {NoStop}%
\bibitem [{\citenamefont {Freese}\ and\ \citenamefont {Miller}(2021)}]{Freese:2021czn}%
  \BibitemOpen
  \bibfield  {author} {\bibinfo {author} {\bibfnamefont {A.}~\bibnamefont {Freese}}\ and\ \bibinfo {author} {\bibfnamefont {G.~A.}\ \bibnamefont {Miller}},\ }\bibfield  {title} {\bibinfo {title} {{Forces within hadrons on the light front}},\ }\href {https://doi.org/10.1103/PhysRevD.103.094023} {\bibfield  {journal} {\bibinfo  {journal} {Phys. Rev. D}\ }\textbf {\bibinfo {volume} {103}},\ \bibinfo {pages} {094023} (\bibinfo {year} {2021})},\ \Eprint {https://arxiv.org/abs/2102.01683} {arXiv:2102.01683 [hep-ph]} \BibitemShut {NoStop}%
\bibitem [{\citenamefont {Freese}(2025)}]{Freese:2025tqd}%
  \BibitemOpen
  \bibfield  {author} {\bibinfo {author} {\bibfnamefont {A.}~\bibnamefont {Freese}},\ }\bibfield  {title} {\bibinfo {title} {{Mechanical form factors and densities of non-relativistic fermions}},\ }\href@noop {} {\  (\bibinfo {year} {2025})},\ \Eprint {https://arxiv.org/abs/2505.06135} {arXiv:2505.06135 [hep-ph]} \BibitemShut {NoStop}%
\bibitem [{\citenamefont {Polyakov}\ and\ \citenamefont {Sun}(2019)}]{Polyakov:2019lbq}%
  \BibitemOpen
  \bibfield  {author} {\bibinfo {author} {\bibfnamefont {M.~V.}\ \bibnamefont {Polyakov}}\ and\ \bibinfo {author} {\bibfnamefont {B.-D.}\ \bibnamefont {Sun}},\ }\bibfield  {title} {\bibinfo {title} {{Gravitational form factors of a spin one particle}},\ }\href {https://doi.org/10.1103/PhysRevD.100.036003} {\bibfield  {journal} {\bibinfo  {journal} {Phys. Rev. D}\ }\textbf {\bibinfo {volume} {100}},\ \bibinfo {pages} {036003} (\bibinfo {year} {2019})},\ \Eprint {https://arxiv.org/abs/1903.02738} {arXiv:1903.02738 [hep-ph]} \BibitemShut {NoStop}%
\bibitem [{\citenamefont {Sun}\ and\ \citenamefont {Dong}(2020)}]{Sun:2020wfo}%
  \BibitemOpen
  \bibfield  {author} {\bibinfo {author} {\bibfnamefont {B.-D.}\ \bibnamefont {Sun}}\ and\ \bibinfo {author} {\bibfnamefont {Y.-B.}\ \bibnamefont {Dong}},\ }\bibfield  {title} {\bibinfo {title} {{Gravitational form factors of $\rho$ meson with a light-cone constituent quark model}},\ }\href {https://doi.org/10.1103/PhysRevD.101.096008} {\bibfield  {journal} {\bibinfo  {journal} {Phys. Rev. D}\ }\textbf {\bibinfo {volume} {101}},\ \bibinfo {pages} {096008} (\bibinfo {year} {2020})},\ \Eprint {https://arxiv.org/abs/2002.02648} {arXiv:2002.02648 [hep-ph]} \BibitemShut {NoStop}%
\bibitem [{\citenamefont {Kim}\ and\ \citenamefont {Sun}(2021)}]{Kim:2020lrs}%
  \BibitemOpen
  \bibfield  {author} {\bibinfo {author} {\bibfnamefont {J.-Y.}\ \bibnamefont {Kim}}\ and\ \bibinfo {author} {\bibfnamefont {B.-D.}\ \bibnamefont {Sun}},\ }\bibfield  {title} {\bibinfo {title} {{Gravitational form factors of a baryon with spin-3/2}},\ }\href {https://doi.org/10.1140/epjc/s10052-021-08852-z} {\bibfield  {journal} {\bibinfo  {journal} {Eur. Phys. J. C}\ }\textbf {\bibinfo {volume} {81}},\ \bibinfo {pages} {85} (\bibinfo {year} {2021})},\ \Eprint {https://arxiv.org/abs/2011.00292} {arXiv:2011.00292 [hep-ph]} \BibitemShut {NoStop}%
\bibitem [{\citenamefont {Barca}\ \emph {et~al.}(2024)\citenamefont {Barca}, \citenamefont {Schaefer}, \citenamefont {Knechtli}, \citenamefont {Urrea-Ni\~no}, \citenamefont {Martins},\ and\ \citenamefont {Peardon}}]{Barca:2024fpc}%
  \BibitemOpen
  \bibfield  {author} {\bibinfo {author} {\bibfnamefont {L.}~\bibnamefont {Barca}}, \bibinfo {author} {\bibfnamefont {S.}~\bibnamefont {Schaefer}}, \bibinfo {author} {\bibfnamefont {F.}~\bibnamefont {Knechtli}}, \bibinfo {author} {\bibfnamefont {J.~A.}\ \bibnamefont {Urrea-Ni\~no}}, \bibinfo {author} {\bibfnamefont {S.}~\bibnamefont {Martins}},\ and\ \bibinfo {author} {\bibfnamefont {M.}~\bibnamefont {Peardon}},\ }\bibfield  {title} {\bibinfo {title} {{Exponential error reduction for glueball calculations using a two-level algorithm in pure gauge theory}},\ }\href {https://doi.org/10.1103/PhysRevD.110.054515} {\bibfield  {journal} {\bibinfo  {journal} {Phys. Rev. D}\ }\textbf {\bibinfo {volume} {110}},\ \bibinfo {pages} {054515} (\bibinfo {year} {2024})},\ \Eprint {https://arxiv.org/abs/2406.12656} {arXiv:2406.12656 [hep-lat]} \BibitemShut {NoStop}%
\bibitem [{\citenamefont {Barca}\ \emph {et~al.}(2025)\citenamefont {Barca}, \citenamefont {Finkenrath}, \citenamefont {Knechtli}, \citenamefont {Peardon}, \citenamefont {Schaefer},\ and\ \citenamefont {Urrea-Ni{\~n}o}}]{Barca:2025fdq}%
  \BibitemOpen
  \bibfield  {author} {\bibinfo {author} {\bibfnamefont {L.}~\bibnamefont {Barca}}, \bibinfo {author} {\bibfnamefont {J.}~\bibnamefont {Finkenrath}}, \bibinfo {author} {\bibfnamefont {F.}~\bibnamefont {Knechtli}}, \bibinfo {author} {\bibfnamefont {M.~J.}\ \bibnamefont {Peardon}}, \bibinfo {author} {\bibfnamefont {S.}~\bibnamefont {Schaefer}},\ and\ \bibinfo {author} {\bibfnamefont {J.~A.}\ \bibnamefont {Urrea-Ni{\~n}o}},\ }\bibfield  {title} {\bibinfo {title} {{Update on two-level sampling for glueball observables in quenched QCD}},\ }\href {https://doi.org/10.22323/1.466.0062} {\bibfield  {journal} {\bibinfo  {journal} {PoS}\ }\textbf {\bibinfo {volume} {LATTICE2024}},\ \bibinfo {pages} {062} (\bibinfo {year} {2025})},\ \Eprint {https://arxiv.org/abs/2501.17988} {arXiv:2501.17988 [hep-lat]} \BibitemShut {NoStop}%
\bibitem [{\citenamefont {Reuther}\ \emph {et~al.}(2018)\citenamefont {Reuther}, \citenamefont {Kepner}, \citenamefont {Byun}, \citenamefont {Samsi}, \citenamefont {Arcand}, \citenamefont {Bestor}, \citenamefont {Bergeron}, \citenamefont {Gadepally}, \citenamefont {Houle}, \citenamefont {Hubbell}, \citenamefont {Jones}, \citenamefont {Klein}, \citenamefont {Milechin}, \citenamefont {Mullen}, \citenamefont {Prout}, \citenamefont {Rosa}, \citenamefont {Yee},\ and\ \citenamefont {Michaleas}}]{reuther2018interactive}%
  \BibitemOpen
  \bibfield  {author} {\bibinfo {author} {\bibfnamefont {A.}~\bibnamefont {Reuther}}, \bibinfo {author} {\bibfnamefont {J.}~\bibnamefont {Kepner}}, \bibinfo {author} {\bibfnamefont {C.}~\bibnamefont {Byun}}, \bibinfo {author} {\bibfnamefont {S.}~\bibnamefont {Samsi}}, \bibinfo {author} {\bibfnamefont {W.}~\bibnamefont {Arcand}}, \bibinfo {author} {\bibfnamefont {D.}~\bibnamefont {Bestor}}, \bibinfo {author} {\bibfnamefont {B.}~\bibnamefont {Bergeron}}, \bibinfo {author} {\bibfnamefont {V.}~\bibnamefont {Gadepally}}, \bibinfo {author} {\bibfnamefont {M.}~\bibnamefont {Houle}}, \bibinfo {author} {\bibfnamefont {M.}~\bibnamefont {Hubbell}}, \bibinfo {author} {\bibfnamefont {M.}~\bibnamefont {Jones}}, \bibinfo {author} {\bibfnamefont {A.}~\bibnamefont {Klein}}, \bibinfo {author} {\bibfnamefont {L.}~\bibnamefont {Milechin}}, \bibinfo {author} {\bibfnamefont {J.}~\bibnamefont {Mullen}}, \bibinfo {author} {\bibfnamefont {A.}~\bibnamefont {Prout}}, \bibinfo {author} {\bibfnamefont {A.}~\bibnamefont {Rosa}}, \bibinfo
  {author} {\bibfnamefont {C.}~\bibnamefont {Yee}},\ and\ \bibinfo {author} {\bibfnamefont {P.}~\bibnamefont {Michaleas}},\ }\bibfield  {title} {\bibinfo {title} {Interactive supercomputing on 40,000 cores for machine learning and data analysis},\ }in\ \href@noop {} {\emph {\bibinfo {booktitle} {2018 IEEE High Performance extreme Computing Conference (HPEC)}}}\ (\bibinfo {organization} {IEEE},\ \bibinfo {year} {2018})\ pp.\ \bibinfo {pages} {1--6}\BibitemShut {NoStop}%
\bibitem [{\citenamefont {Bendavid}\ \emph {et~al.}(2025)\citenamefont {Bendavid}, \citenamefont {D'Alfonso}, \citenamefont {Eysermans}, \citenamefont {Freer}, \citenamefont {Goncharov}, \citenamefont {Heine}, \citenamefont {Lavezzo}, \citenamefont {Moore}, \citenamefont {Paus}, \citenamefont {Shen}, \citenamefont {Walter},\ and\ \citenamefont {Wang}}]{bendavid2025submitphysicsanalysisfacility}%
  \BibitemOpen
  \bibfield  {author} {\bibinfo {author} {\bibfnamefont {J.}~\bibnamefont {Bendavid}}, \bibinfo {author} {\bibfnamefont {M.}~\bibnamefont {D'Alfonso}}, \bibinfo {author} {\bibfnamefont {J.}~\bibnamefont {Eysermans}}, \bibinfo {author} {\bibfnamefont {C.}~\bibnamefont {Freer}}, \bibinfo {author} {\bibfnamefont {M.}~\bibnamefont {Goncharov}}, \bibinfo {author} {\bibfnamefont {M.}~\bibnamefont {Heine}}, \bibinfo {author} {\bibfnamefont {L.}~\bibnamefont {Lavezzo}}, \bibinfo {author} {\bibfnamefont {M.}~\bibnamefont {Moore}}, \bibinfo {author} {\bibfnamefont {C.}~\bibnamefont {Paus}}, \bibinfo {author} {\bibfnamefont {X.}~\bibnamefont {Shen}}, \bibinfo {author} {\bibfnamefont {D.}~\bibnamefont {Walter}},\ and\ \bibinfo {author} {\bibfnamefont {Z.}~\bibnamefont {Wang}},\ }\href {https://arxiv.org/abs/2506.01958} {\bibinfo {title} {Submit: A physics analysis facility at mit}} (\bibinfo {year} {2025}),\ \Eprint {https://arxiv.org/abs/2506.01958} {arXiv:2506.01958 [cs.DC]} \BibitemShut {NoStop}%
\bibitem [{\citenamefont {Edwards}\ and\ \citenamefont {Joó}(2005)}]{Edwards:2004sx}%
  \BibitemOpen
  \bibfield  {author} {\bibinfo {author} {\bibfnamefont {R.~G.}\ \bibnamefont {Edwards}}\ and\ \bibinfo {author} {\bibfnamefont {B.}~\bibnamefont {Joó}} (\bibinfo {collaboration} {SciDAC Collaboration, LHPC Collaboration, UKQCD Collaboration}),\ }\bibfield  {title} {\bibinfo {title} {{The Chroma software system for lattice QCD}},\ }\href {https://doi.org/10.1016/j.nuclphysbps.2004.11.254} {\bibfield  {journal} {\bibinfo  {journal} {Nucl.Phys.Proc.Suppl.}\ }\textbf {\bibinfo {volume} {140}},\ \bibinfo {pages} {832} (\bibinfo {year} {2005})},\ \Eprint {https://arxiv.org/abs/hep-lat/0409003} {arXiv:hep-lat/0409003 [hep-lat]} \BibitemShut {NoStop}%
%%CITATION = HEP-LAT/0409003;%%
\bibitem [{\citenamefont {Gambhir}\ \emph {et~al.}(2018)\citenamefont {Gambhir}, \citenamefont {Brantley}, \citenamefont {Chang}, \citenamefont {Hörz}, \citenamefont {Monge-Camacho}, \citenamefont {Vranas},\ and\ \citenamefont {Walker-Loud}}]{lalibe}%
  \BibitemOpen
  \bibfield  {author} {\bibinfo {author} {\bibfnamefont {A.}~\bibnamefont {Gambhir}}, \bibinfo {author} {\bibfnamefont {D.}~\bibnamefont {Brantley}}, \bibinfo {author} {\bibfnamefont {J.}~\bibnamefont {Chang}}, \bibinfo {author} {\bibfnamefont {B.}~\bibnamefont {Hörz}}, \bibinfo {author} {\bibfnamefont {H.}~\bibnamefont {Monge-Camacho}}, \bibinfo {author} {\bibfnamefont {P.}~\bibnamefont {Vranas}},\ and\ \bibinfo {author} {\bibfnamefont {A.}~\bibnamefont {Walker-Loud}},\ }\href@noop {} {\bibinfo {title} {lalibe}},\ \bibinfo {howpublished} {\url{https://github.com/callat-qcd/lalibe}} (\bibinfo {year} {2018})\BibitemShut {NoStop}%
\bibitem [{\citenamefont {Bradbury}\ \emph {et~al.}(2018)\citenamefont {Bradbury}, \citenamefont {Frostig}, \citenamefont {Hawkins}, \citenamefont {Johnson}, \citenamefont {Leary}, \citenamefont {Maclaurin}, \citenamefont {Necula}, \citenamefont {Paszke}, \citenamefont {Vander{P}las}, \citenamefont {Wanderman-{M}ilne},\ and\ \citenamefont {Zhang}}]{jax2018github}%
  \BibitemOpen
  \bibfield  {author} {\bibinfo {author} {\bibfnamefont {J.}~\bibnamefont {Bradbury}}, \bibinfo {author} {\bibfnamefont {R.}~\bibnamefont {Frostig}}, \bibinfo {author} {\bibfnamefont {P.}~\bibnamefont {Hawkins}}, \bibinfo {author} {\bibfnamefont {M.~J.}\ \bibnamefont {Johnson}}, \bibinfo {author} {\bibfnamefont {C.}~\bibnamefont {Leary}}, \bibinfo {author} {\bibfnamefont {D.}~\bibnamefont {Maclaurin}}, \bibinfo {author} {\bibfnamefont {G.}~\bibnamefont {Necula}}, \bibinfo {author} {\bibfnamefont {A.}~\bibnamefont {Paszke}}, \bibinfo {author} {\bibfnamefont {J.}~\bibnamefont {Vander{P}las}}, \bibinfo {author} {\bibfnamefont {S.}~\bibnamefont {Wanderman-{M}ilne}},\ and\ \bibinfo {author} {\bibfnamefont {Q.}~\bibnamefont {Zhang}},\ }\href {http://github.com/jax-ml/jax} {\bibinfo {title} {{JAX}: composable transformations of {P}ython+{N}um{P}y programs}} (\bibinfo {year} {2018})\BibitemShut {NoStop}%
\bibitem [{\citenamefont {Harris}\ \emph {et~al.}(2020)\citenamefont {Harris}, \citenamefont {Millman}, \citenamefont {van~der Walt}, \citenamefont {Gommers}, \citenamefont {Virtanen}, \citenamefont {Cournapeau}, \citenamefont {Wieser}, \citenamefont {Taylor}, \citenamefont {Berg}, \citenamefont {Smith}, \citenamefont {Kern}, \citenamefont {Picus}, \citenamefont {Hoyer}, \citenamefont {van Kerkwijk}, \citenamefont {Brett}, \citenamefont {Haldane}, \citenamefont {del R{\'{i}}o}, \citenamefont {Wiebe}, \citenamefont {Peterson}, \citenamefont {G{\'{e}}rard-Marchant}, \citenamefont {Sheppard}, \citenamefont {Reddy}, \citenamefont {Weckesser}, \citenamefont {Abbasi}, \citenamefont {Gohlke},\ and\ \citenamefont {Oliphant}}]{harris2020array}%
  \BibitemOpen
  \bibfield  {author} {\bibinfo {author} {\bibfnamefont {C.~R.}\ \bibnamefont {Harris}}, \bibinfo {author} {\bibfnamefont {K.~J.}\ \bibnamefont {Millman}}, \bibinfo {author} {\bibfnamefont {S.~J.}\ \bibnamefont {van~der Walt}}, \bibinfo {author} {\bibfnamefont {R.}~\bibnamefont {Gommers}}, \bibinfo {author} {\bibfnamefont {P.}~\bibnamefont {Virtanen}}, \bibinfo {author} {\bibfnamefont {D.}~\bibnamefont {Cournapeau}}, \bibinfo {author} {\bibfnamefont {E.}~\bibnamefont {Wieser}}, \bibinfo {author} {\bibfnamefont {J.}~\bibnamefont {Taylor}}, \bibinfo {author} {\bibfnamefont {S.}~\bibnamefont {Berg}}, \bibinfo {author} {\bibfnamefont {N.~J.}\ \bibnamefont {Smith}}, \bibinfo {author} {\bibfnamefont {R.}~\bibnamefont {Kern}}, \bibinfo {author} {\bibfnamefont {M.}~\bibnamefont {Picus}}, \bibinfo {author} {\bibfnamefont {S.}~\bibnamefont {Hoyer}}, \bibinfo {author} {\bibfnamefont {M.~H.}\ \bibnamefont {van Kerkwijk}}, \bibinfo {author} {\bibfnamefont {M.}~\bibnamefont {Brett}}, \bibinfo {author} {\bibfnamefont
  {A.}~\bibnamefont {Haldane}}, \bibinfo {author} {\bibfnamefont {J.~F.}\ \bibnamefont {del R{\'{i}}o}}, \bibinfo {author} {\bibfnamefont {M.}~\bibnamefont {Wiebe}}, \bibinfo {author} {\bibfnamefont {P.}~\bibnamefont {Peterson}}, \bibinfo {author} {\bibfnamefont {P.}~\bibnamefont {G{\'{e}}rard-Marchant}}, \bibinfo {author} {\bibfnamefont {K.}~\bibnamefont {Sheppard}}, \bibinfo {author} {\bibfnamefont {T.}~\bibnamefont {Reddy}}, \bibinfo {author} {\bibfnamefont {W.}~\bibnamefont {Weckesser}}, \bibinfo {author} {\bibfnamefont {H.}~\bibnamefont {Abbasi}}, \bibinfo {author} {\bibfnamefont {C.}~\bibnamefont {Gohlke}},\ and\ \bibinfo {author} {\bibfnamefont {T.~E.}\ \bibnamefont {Oliphant}},\ }\bibfield  {title} {\bibinfo {title} {Array programming with {NumPy}},\ }\href {https://doi.org/10.1038/s41586-020-2649-2} {\bibfield  {journal} {\bibinfo  {journal} {Nature}\ }\textbf {\bibinfo {volume} {585}},\ \bibinfo {pages} {357} (\bibinfo {year} {2020})}\BibitemShut {NoStop}%
\bibitem [{\citenamefont {Virtanen}\ \emph {et~al.}(2020)\citenamefont {Virtanen}, \citenamefont {Gommers}, \citenamefont {Oliphant}, \citenamefont {Haberland}, \citenamefont {Reddy}, \citenamefont {Cournapeau}, \citenamefont {Burovski}, \citenamefont {Peterson}, \citenamefont {Weckesser}, \citenamefont {Bright}, \citenamefont {{van der Walt}}, \citenamefont {Brett}, \citenamefont {Wilson}, \citenamefont {Millman}, \citenamefont {Mayorov}, \citenamefont {Nelson}, \citenamefont {Jones}, \citenamefont {Kern}, \citenamefont {Larson}, \citenamefont {Carey}, \citenamefont {Polat}, \citenamefont {Feng}, \citenamefont {Moore}, \citenamefont {{VanderPlas}}, \citenamefont {Laxalde}, \citenamefont {Perktold}, \citenamefont {Cimrman}, \citenamefont {Henriksen}, \citenamefont {Quintero}, \citenamefont {Harris}, \citenamefont {Archibald}, \citenamefont {Ribeiro}, \citenamefont {Pedregosa}, \citenamefont {{van Mulbregt}},\ and\ \citenamefont {{SciPy 1.0 Contributors}}}]{2020SciPy-NMeth}%
  \BibitemOpen
  \bibfield  {author} {\bibinfo {author} {\bibfnamefont {P.}~\bibnamefont {Virtanen}}, \bibinfo {author} {\bibfnamefont {R.}~\bibnamefont {Gommers}}, \bibinfo {author} {\bibfnamefont {T.~E.}\ \bibnamefont {Oliphant}}, \bibinfo {author} {\bibfnamefont {M.}~\bibnamefont {Haberland}}, \bibinfo {author} {\bibfnamefont {T.}~\bibnamefont {Reddy}}, \bibinfo {author} {\bibfnamefont {D.}~\bibnamefont {Cournapeau}}, \bibinfo {author} {\bibfnamefont {E.}~\bibnamefont {Burovski}}, \bibinfo {author} {\bibfnamefont {P.}~\bibnamefont {Peterson}}, \bibinfo {author} {\bibfnamefont {W.}~\bibnamefont {Weckesser}}, \bibinfo {author} {\bibfnamefont {J.}~\bibnamefont {Bright}}, \bibinfo {author} {\bibfnamefont {S.~J.}\ \bibnamefont {{van der Walt}}}, \bibinfo {author} {\bibfnamefont {M.}~\bibnamefont {Brett}}, \bibinfo {author} {\bibfnamefont {J.}~\bibnamefont {Wilson}}, \bibinfo {author} {\bibfnamefont {K.~J.}\ \bibnamefont {Millman}}, \bibinfo {author} {\bibfnamefont {N.}~\bibnamefont {Mayorov}}, \bibinfo {author} {\bibfnamefont
  {A.~R.~J.}\ \bibnamefont {Nelson}}, \bibinfo {author} {\bibfnamefont {E.}~\bibnamefont {Jones}}, \bibinfo {author} {\bibfnamefont {R.}~\bibnamefont {Kern}}, \bibinfo {author} {\bibfnamefont {E.}~\bibnamefont {Larson}}, \bibinfo {author} {\bibfnamefont {C.~J.}\ \bibnamefont {Carey}}, \bibinfo {author} {\bibfnamefont {{\.I}.}~\bibnamefont {Polat}}, \bibinfo {author} {\bibfnamefont {Y.}~\bibnamefont {Feng}}, \bibinfo {author} {\bibfnamefont {E.~W.}\ \bibnamefont {Moore}}, \bibinfo {author} {\bibfnamefont {J.}~\bibnamefont {{VanderPlas}}}, \bibinfo {author} {\bibfnamefont {D.}~\bibnamefont {Laxalde}}, \bibinfo {author} {\bibfnamefont {J.}~\bibnamefont {Perktold}}, \bibinfo {author} {\bibfnamefont {R.}~\bibnamefont {Cimrman}}, \bibinfo {author} {\bibfnamefont {I.}~\bibnamefont {Henriksen}}, \bibinfo {author} {\bibfnamefont {E.~A.}\ \bibnamefont {Quintero}}, \bibinfo {author} {\bibfnamefont {C.~R.}\ \bibnamefont {Harris}}, \bibinfo {author} {\bibfnamefont {A.~M.}\ \bibnamefont {Archibald}}, \bibinfo {author}
  {\bibfnamefont {A.~H.}\ \bibnamefont {Ribeiro}}, \bibinfo {author} {\bibfnamefont {F.}~\bibnamefont {Pedregosa}}, \bibinfo {author} {\bibfnamefont {P.}~\bibnamefont {{van Mulbregt}}},\ and\ \bibinfo {author} {\bibnamefont {{SciPy 1.0 Contributors}}},\ }\bibfield  {title} {\bibinfo {title} {{{SciPy} 1.0: Fundamental Algorithms for Scientific Computing in Python}},\ }\href {https://doi.org/10.1038/s41592-019-0686-2} {\bibfield  {journal} {\bibinfo  {journal} {Nature Methods}\ }\textbf {\bibinfo {volume} {17}},\ \bibinfo {pages} {261} (\bibinfo {year} {2020})}\BibitemShut {NoStop}%
\bibitem [{\citenamefont {Lepage}(2020{\natexlab{a}})}]{peter_lepage_2020_4037174}%
  \BibitemOpen
  \bibfield  {author} {\bibinfo {author} {\bibfnamefont {G.~P.}\ \bibnamefont {Lepage}},\ }\bibfield  {title} {\bibinfo {title} {{lsqfit v. 11.7}}\ }\href {https://doi.org/doi:10.5281/zenodo.4037174} {doi:10.5281/zenodo.4037174} (\bibinfo {year} {2020}{\natexlab{a}}),\ \Eprint {https://arxiv.org/abs/https://github.com/gplepage/lsqfit} {https://github.com/gplepage/lsqfit} \BibitemShut {NoStop}%
\bibitem [{\citenamefont {Lepage}(2020{\natexlab{b}})}]{peter_lepage_2020_4290884}%
  \BibitemOpen
  \bibfield  {author} {\bibinfo {author} {\bibfnamefont {G.~P.}\ \bibnamefont {Lepage}},\ }\bibfield  {title} {\bibinfo {title} {{gvar v. 11.9.1}}\ }\href {https://doi.org/doi:10.5281/zenodo.4290884} {doi:10.5281/zenodo.4290884} (\bibinfo {year} {2020}{\natexlab{b}}),\ \Eprint {https://arxiv.org/abs/https://github.com/gplepage/gvar} {https://github.com/gplepage/gvar} \BibitemShut {NoStop}%
\bibitem [{\citenamefont {Hunter}(2007)}]{Hunter:2007}%
  \BibitemOpen
  \bibfield  {author} {\bibinfo {author} {\bibfnamefont {J.~D.}\ \bibnamefont {Hunter}},\ }\bibfield  {title} {\bibinfo {title} {Matplotlib: A 2d graphics environment},\ }\href {https://doi.org/10.1109/MCSE.2007.55} {\bibfield  {journal} {\bibinfo  {journal} {Computing in Science \& Engineering}\ }\textbf {\bibinfo {volume} {9}},\ \bibinfo {pages} {90} (\bibinfo {year} {2007})}\BibitemShut {NoStop}%
\bibitem [{\citenamefont {Jay}\ and\ \citenamefont {Neil}(2021)}]{Jay:2020jkz}%
  \BibitemOpen
  \bibfield  {author} {\bibinfo {author} {\bibfnamefont {W.~I.}\ \bibnamefont {Jay}}\ and\ \bibinfo {author} {\bibfnamefont {E.~T.}\ \bibnamefont {Neil}},\ }\bibfield  {title} {\bibinfo {title} {{Bayesian model averaging for analysis of lattice field theory results}},\ }\href {https://doi.org/10.1103/PhysRevD.103.114502} {\bibfield  {journal} {\bibinfo  {journal} {Phys. Rev. D}\ }\textbf {\bibinfo {volume} {103}},\ \bibinfo {pages} {114502} (\bibinfo {year} {2021})},\ \Eprint {https://arxiv.org/abs/2008.01069} {arXiv:2008.01069 [stat.ME]} \BibitemShut {NoStop}%
\bibitem [{\citenamefont {Rinaldi}\ \emph {et~al.}(2019)\citenamefont {Rinaldi}, \citenamefont {Syritsyn}, \citenamefont {Wagman}, \citenamefont {Buchoff}, \citenamefont {Schroeder},\ and\ \citenamefont {Wasem}}]{Rinaldi:2018osy}%
  \BibitemOpen
  \bibfield  {author} {\bibinfo {author} {\bibfnamefont {E.}~\bibnamefont {Rinaldi}}, \bibinfo {author} {\bibfnamefont {S.}~\bibnamefont {Syritsyn}}, \bibinfo {author} {\bibfnamefont {M.~L.}\ \bibnamefont {Wagman}}, \bibinfo {author} {\bibfnamefont {M.~I.}\ \bibnamefont {Buchoff}}, \bibinfo {author} {\bibfnamefont {C.}~\bibnamefont {Schroeder}},\ and\ \bibinfo {author} {\bibfnamefont {J.}~\bibnamefont {Wasem}},\ }\bibfield  {title} {\bibinfo {title} {{Neutron-antineutron oscillations from lattice QCD}},\ }\href {https://doi.org/10.1103/PhysRevLett.122.162001} {\bibfield  {journal} {\bibinfo  {journal} {Phys. Rev. Lett.}\ }\textbf {\bibinfo {volume} {122}},\ \bibinfo {pages} {162001} (\bibinfo {year} {2019})},\ \Eprint {https://arxiv.org/abs/1809.00246} {arXiv:1809.00246 [hep-lat]} \BibitemShut {NoStop}%
\bibitem [{\citenamefont {Beane}\ \emph {et~al.}(2015)\citenamefont {Beane}, \citenamefont {Chang}, \citenamefont {Detmold}, \citenamefont {Orginos}, \citenamefont {Parre{\~n}o}, \citenamefont {Savage},\ and\ \citenamefont {Tiburzi}}]{Beane:2015yha}%
  \BibitemOpen
  \bibfield  {author} {\bibinfo {author} {\bibfnamefont {S.~R.}\ \bibnamefont {Beane}}, \bibinfo {author} {\bibfnamefont {E.}~\bibnamefont {Chang}}, \bibinfo {author} {\bibfnamefont {W.}~\bibnamefont {Detmold}}, \bibinfo {author} {\bibfnamefont {K.}~\bibnamefont {Orginos}}, \bibinfo {author} {\bibfnamefont {A.}~\bibnamefont {Parre{\~n}o}}, \bibinfo {author} {\bibfnamefont {M.~J.}\ \bibnamefont {Savage}},\ and\ \bibinfo {author} {\bibfnamefont {B.~C.}\ \bibnamefont {Tiburzi}} (\bibinfo {collaboration} {NPLQCD}),\ }\bibfield  {title} {\bibinfo {title} {{Ab initio Calculation of the np{\textrightarrow}d{\ensuremath{\gamma}} Radiative Capture Process}},\ }\href {https://doi.org/10.1103/PhysRevLett.115.132001} {\bibfield  {journal} {\bibinfo  {journal} {Phys. Rev. Lett.}\ }\textbf {\bibinfo {volume} {115}},\ \bibinfo {pages} {132001} (\bibinfo {year} {2015})},\ \Eprint {https://arxiv.org/abs/1505.02422} {arXiv:1505.02422 [hep-lat]} \BibitemShut {NoStop}%
\bibitem [{\citenamefont {Luscher}\ and\ \citenamefont {Weisz}(1985)}]{Luscher:1984xn}%
  \BibitemOpen
  \bibfield  {author} {\bibinfo {author} {\bibfnamefont {M.}~\bibnamefont {Luscher}}\ and\ \bibinfo {author} {\bibfnamefont {P.}~\bibnamefont {Weisz}},\ }\bibfield  {title} {\bibinfo {title} {{On-shell improved lattice gauge theories}},\ }\href {https://doi.org/10.1007/BF01205792} {\bibfield  {journal} {\bibinfo  {journal} {Commun. Math. Phys.}\ }\textbf {\bibinfo {volume} {98}},\ \bibinfo {pages} {433} (\bibinfo {year} {1985})},\ \bibinfo {note} {[Erratum: Commun.Math.Phys. 98, 433 (1985)]}\BibitemShut {NoStop}%
\bibitem [{\citenamefont {Sheikholeslami}\ and\ \citenamefont {Wohlert}(1985)}]{Sheikholeslami:1985ij}%
  \BibitemOpen
  \bibfield  {author} {\bibinfo {author} {\bibfnamefont {B.}~\bibnamefont {Sheikholeslami}}\ and\ \bibinfo {author} {\bibfnamefont {R.}~\bibnamefont {Wohlert}},\ }\bibfield  {title} {\bibinfo {title} {{Improved Continuum Limit Lattice Action for QCD with Wilson Fermions}},\ }\href {https://doi.org/10.1016/0550-3213(85)90002-1} {\bibfield  {journal} {\bibinfo  {journal} {Nucl. Phys. B}\ }\textbf {\bibinfo {volume} {259}},\ \bibinfo {pages} {572} (\bibinfo {year} {1985})}\BibitemShut {NoStop}%
\end{thebibliography}%

\newpage

\onecolumngrid

\section{Additional results}

\begin{figure*}
\includegraphics[width=0.98\linewidth]{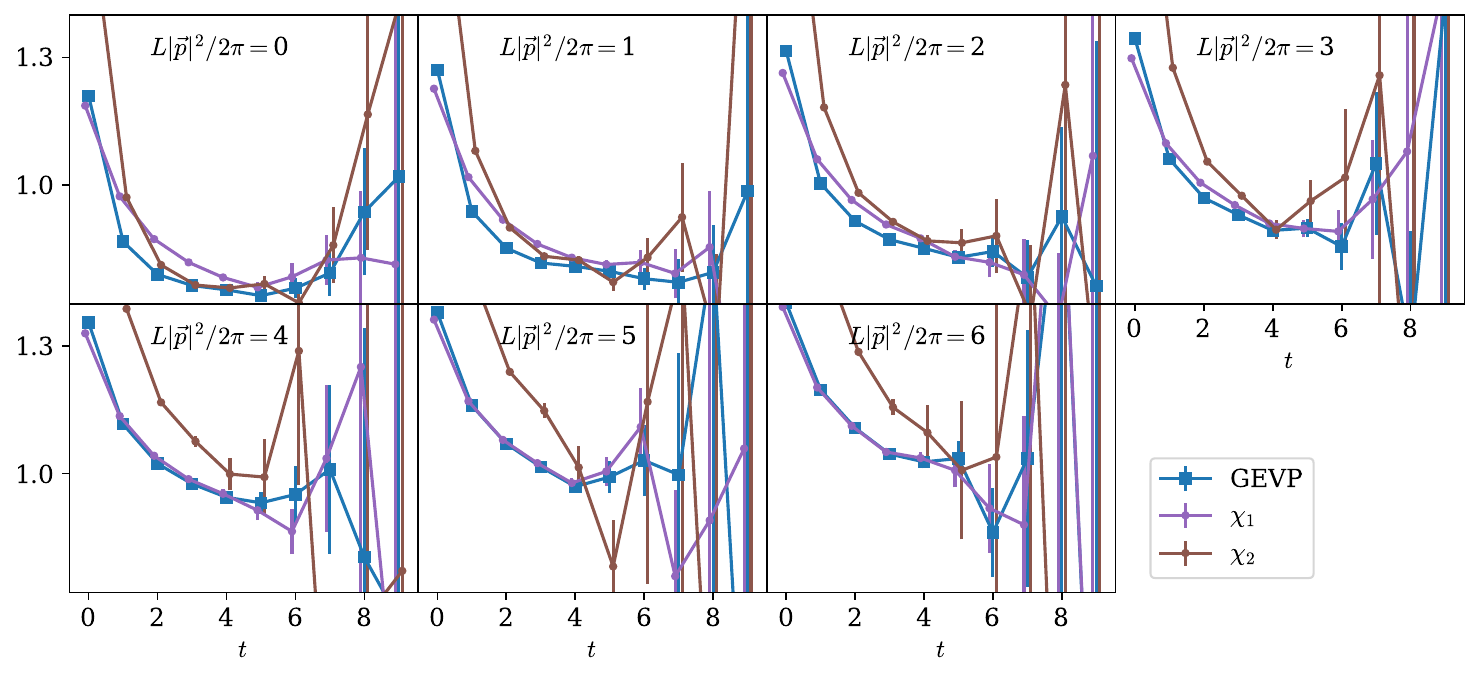}
\caption{
    Effective energies $E_{\text{eff}}(t)=\text{log}(C^{2\text{pt}}(t)/C^{2\text{pt}}(t+1))$ as a function of sink time $t$ for the different boost momenta, comparing the diagonal correlators $C^{2\text{pt}}_{ii}$ of Eq.~\eqref{eq:2pt-matrix} computed with single interpolators from the basis Eq.~\eqref{eq:interp} with the GEVP-optimized one of Eq.~\eqref{eq:gevp-2pt} defined with the composite interpolator Eq.~\eqref{eq:interp-gevp}
}
\label{fig:spectrum}
\end{figure*}

{\bf GEVP analysis}: Figure~\ref{fig:spectrum} compares the ground-state energies extracted using GEVP-optimized interpolators with the individual interpolators in the basis.

\begin{figure}
\includegraphics[width=0.98\linewidth]{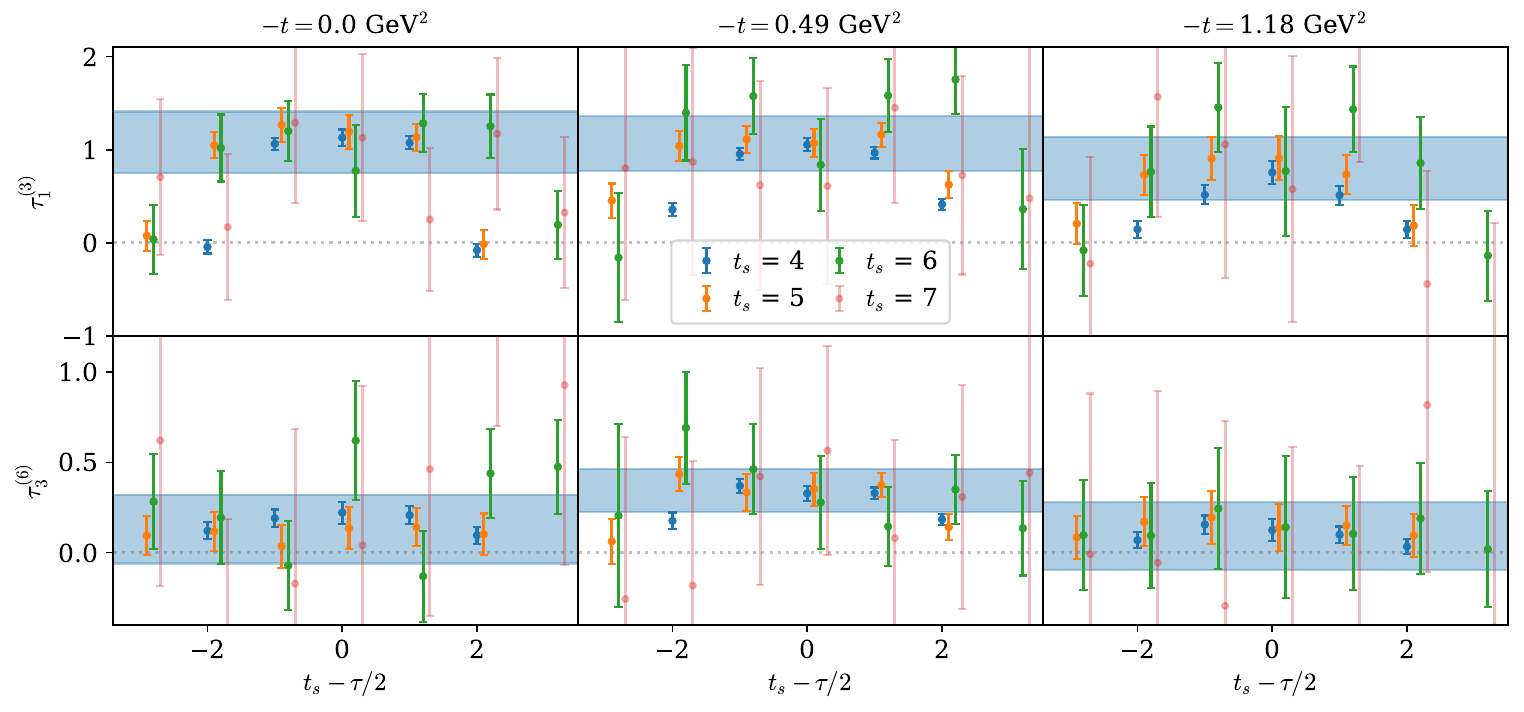}
\caption{Example averaged ratios of three- and two-point functions at different sink times $t_s$, expected to be proportional to the matrix elements of Eq.~\eqref{eq:ME} up to known kinematic factors. Deviation from a constant value (i.e., curvature) is an indication of excited-state contamination effects. The bands correspond to model averages over fits to different time ranges. The two rows show examples from each of the two irreps, and the three columns to different momentum transfers $t$.}
\label{fig:ME}
\end{figure}

{\bf Matrix element fits}: The matrix elements are extracted via a constant fit to all possible ($t_s$,$\tau$) time ranges of the ratios of Eq.~\eqref{eq:ratios}, after having averaged over all kinematics and irrep vectors that correspond to identical linear combinations of GFFs. The minimum number of data points per fit is set to 8, and additionally $t_{s_{\text{min}}} \geq 5$, $t_{s_{\text{max}}} \leq 13$, $\tau \geq 2$,  $\tau \leq t_s - 2$ are imposed. The final matrix element results are averages of all fits weighted by their $p$-value weights~\cite{Jay:2020jkz,Rinaldi:2018osy,Beane:2015yha}. Examples of ratios and corresponding fits are shown in Fig~\ref{fig:ME}.

\begin{figure}
\includegraphics[width=0.98\linewidth]{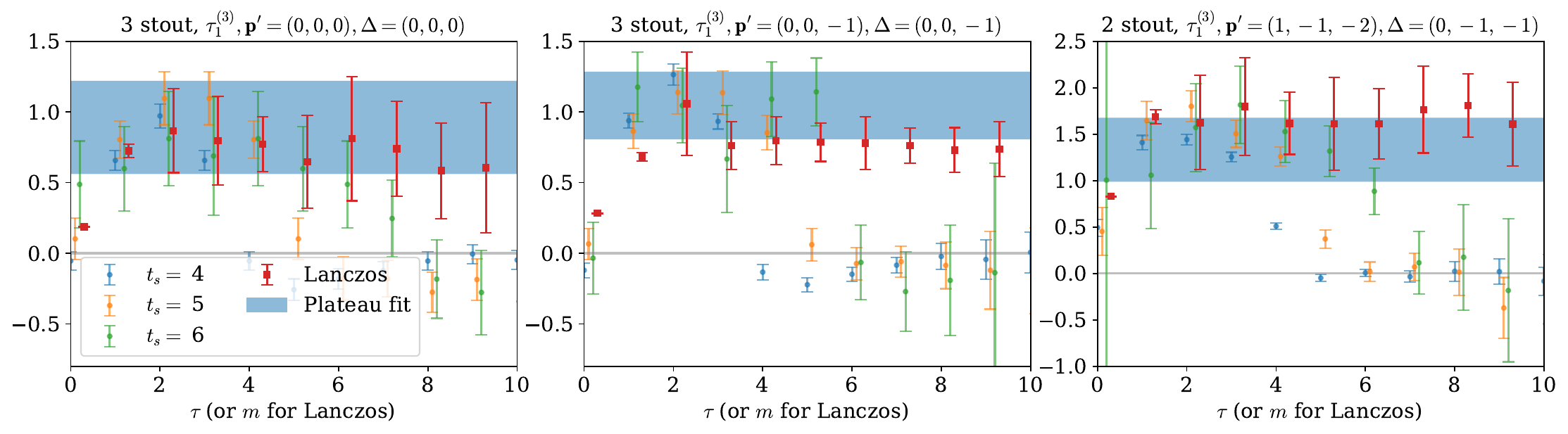}
\caption{
    Comparison between the ratio fitting procedure and a Lanczos analysis, for three different three-point functions with definite kinematics as indicated.
    The data are shown as ratios as elsewhere, except plotted here versus $\tau$.
    Lanczos results are shown versus iteration $m$, which is not directly equivalent to $\tau$; the data are plotted together to allow visual comparison of the plateau value.
}
\label{fig:Lanczos}
\end{figure}

Fig.~\ref{fig:Lanczos} shows the results of a cross-check of our fitting procedure against a block Lanczos matrix element analysis~\cite{Wagman:2024rid,Hackett:2024xnx,Hackett:2024nbe,Abbott:2025yhm}.
The Lanczos approach requires a three-point function and the corresponding initial- and final-state two-point functions as inputs, and so is inapplicable to the binned ratios fit for the primary analysis.
Data were not available for a full reanalysis using Lanczos methods.
Thus, for the test, we repeat the ratio fitting procedure on a small subset of unbinned ratios with definite kinematics.
For the Lanczos analysis of the corresponding two- and three-point functions, we use Hermitian subspace~\cite{Hackett:2024xnx} and ZCW~\cite{Hackett:2024nbe} filtering with $\epsilon^\mathrm{ZCW} = 10^{-3}$, and use a nested bootstrap median estimator to compute values and errors.
The resulting estimates shown in Fig~\ref{fig:Lanczos} converge to a stable value after a few iterations.
In all cases we have examined, the value provided is statistically consistent with the fit result; the uncertainties on fit and Lanczos estimates are comparable.

\begin{figure*}
\includegraphics[width=0.98\linewidth]{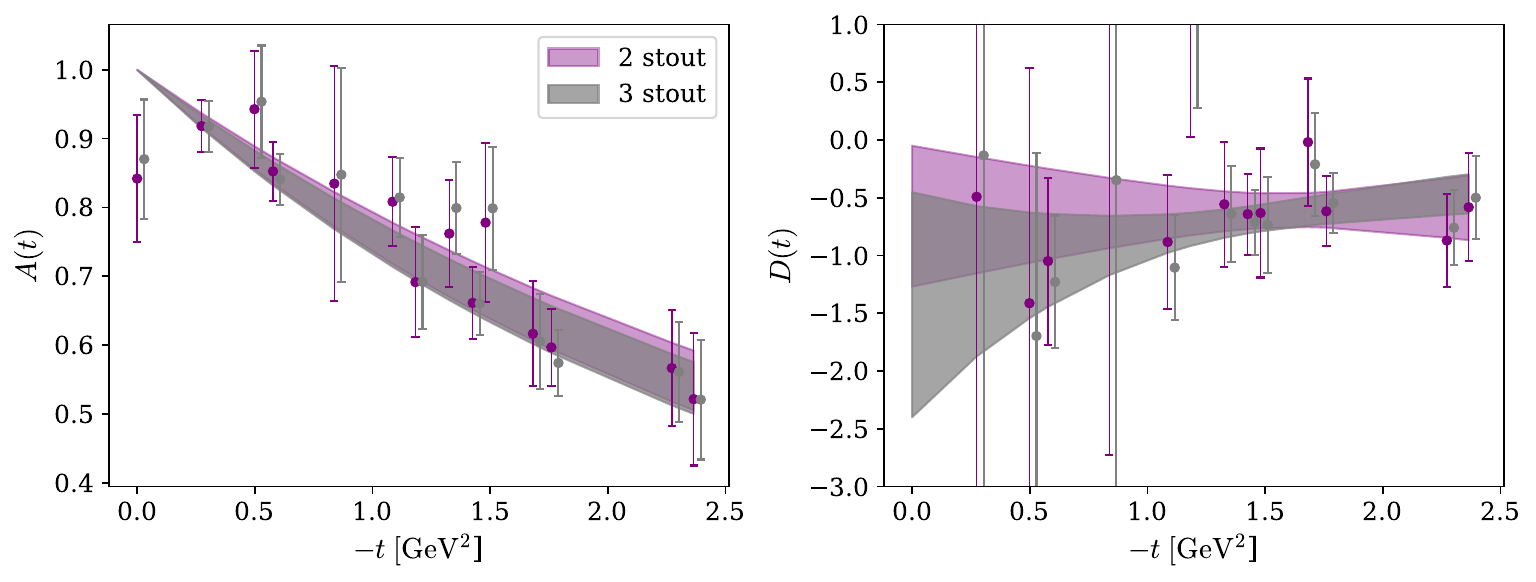}
\caption{
Comparison between the results obtained using a gluon EMT with two and three steps of stout smearing, renormalized such that $A(0)=1$. The bands are tripole ($n=3$) fits to the GFFs. The purple markers have been slightly shifted to the right for visibility.
}
\label{fig:tt_ttt}
\end{figure*}

\begin{figure*}
\includegraphics[width=0.98\linewidth]{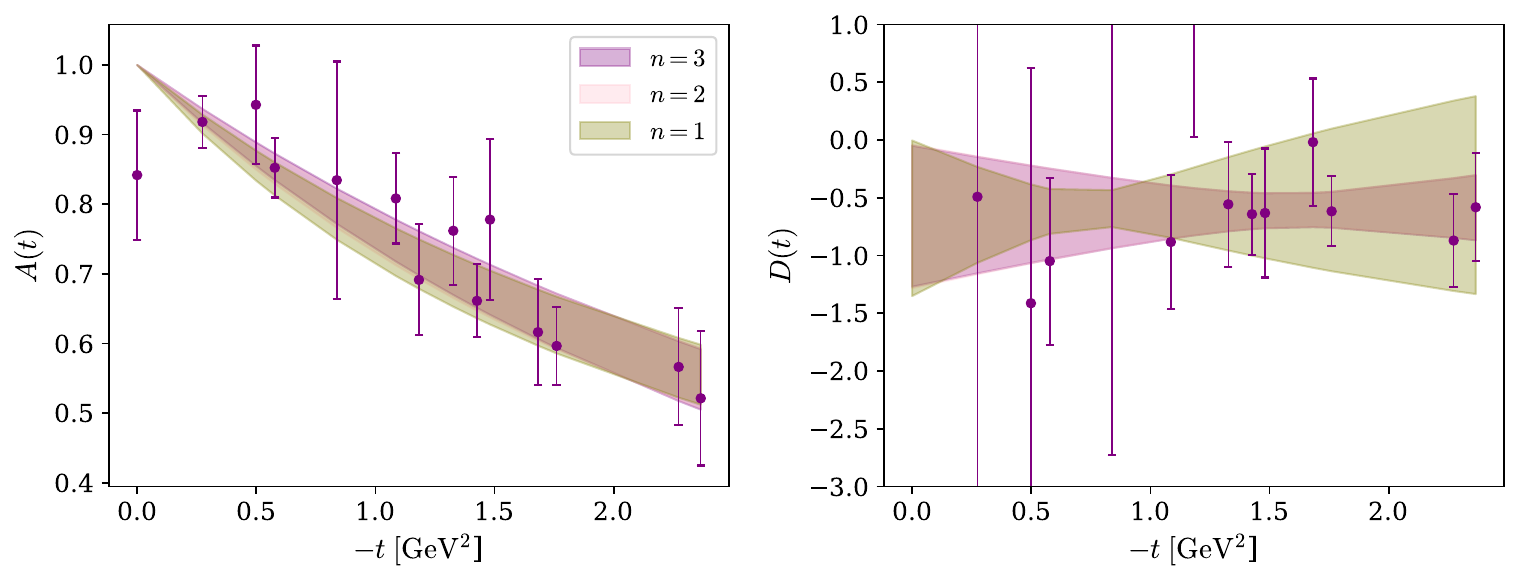}
\caption{
Comparison between multipole fits with $n=\{1,2,3\}$ to the results obtained using a gluon EMT with three steps of stout smearing, renormalized such that $A(0)=1$.
}
\label{fig:multipoles}
\end{figure*}

{\bf Additional GFF results}: Figure~\ref{fig:tt_ttt} shows a comparison of the GFFs using 3 versus 2 steps of stout smearing for the EMT. In Fig.~\ref{fig:multipoles}, we compare different choices of the $n$-pole model ($n=1,2,3$) for the 2-stout smeared operator.

\section{Pseudoscalar meson analysis}
\label{app:B}

Figure~\ref{fig:GFFs} of the main text presents a comparison of the results of this work against a previous extraction of the gluon GFFs of four hadrons of different spin~\cite{Pefkou:2021fni}. The older calculation was performed with a L\"uscher-Weisz gauge action \cite{Luscher:1984xn}, and $N_f=2+1$ flavors of non-perturbatively improved clover Wilson quarks \cite{Sheikholeslami:1985ij}, with their bare masses tuned yield a pion mass of $m_{\pi}\simeq 450~\text{MeV}$. Since for the glueball in Yang-Mills the overall normalization of $A_g(t)$ is trivial and set as $A_g(0)=1$, the glueball GFFs are compared with $A_g(t)/A_g(0)$ and $D_g(t)/A_g(0)$ of the other hadrons. However, given the difference in actions used between the two calculations, it is important to assess to what extent the comparison of the rest of the properties---the $t$-dependence of $A(t)$ and $D(t)$, as well as $D(0)/A(0)$---is physically meaningful. For that purpose, a quenched calculation of the pion gluon GFFs is undertaken using the same Yang-Mills action as was used for the glueball calculation, and clover-improved Wilson valence quarks with clover coefficient set to 1, and bare mass tuned to match the pion mass of Ref.~\cite{Pefkou:2021fni}, $m_{\pi}\simeq 450~\text{MeV}$. Mixing between gluon and quark GFFs is neglected, as was done in Ref.~\cite{Pefkou:2021fni}.

A single interpolating operator $\overline{u}(x)\gamma_5 d(x)$ is used, with quark fields that have been smeared
by Gaussian gauge-invariant smearing
with radius $4.7a$. The two-point correlator is measured at 12 spacetime sources of 2760 configurations, and momenta $|\vec{p}|^2\leq 6(2\pi/L)^2$. A GEVP analysis using a $2\times2$ matrix of two-point correlation functions constructed using both smeared-smeared and point-smeared propagators yields consistent results as the non-GEVP smeared-smeared only analysis, so the latter is taken as the primary result. The two-point correlator is multiplied with the EMT operator on each of the 2760 configurations, using an EMT with 2 and 3 steps of stout smearing and momenta $\left|\vec{\Delta}\right|^2 \leq 10 (2\pi/L)^2$. 200 bootstrap ensembles of vacuum substracted three-point functions are taken to form the ratios of Eq.~\eqref{eq:ratios}. The rest of the analysis proceeds identically to the glueball GFF extraction.

\begin{figure*} 
\includegraphics[width=0.98\linewidth]{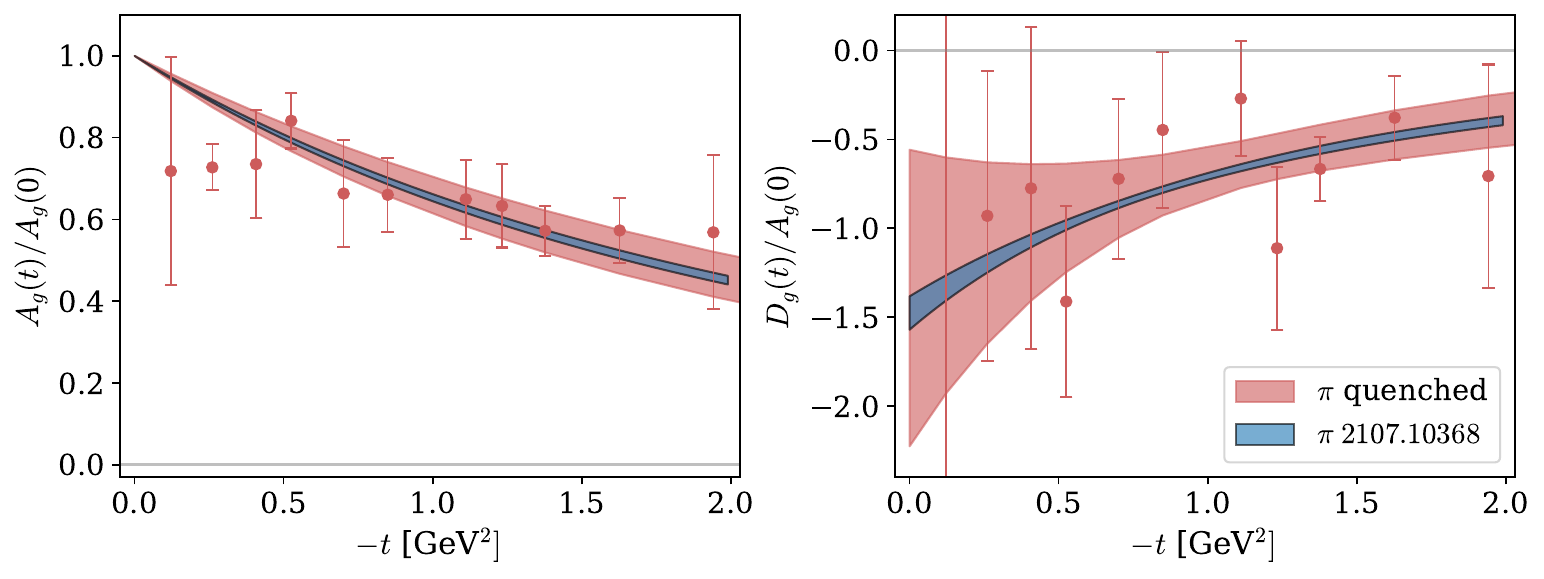}
\caption{
Comparison between the pseudoscalar meson gluon GFFs in the quenched theory obtained with the ensemble used in this work, and the gluon GFFs of 
the pion obtained 
with an $N_f=2+1$ QCD ensemble with $m_{\pi}\simeq 450~\text{MeV}$~\cite{Pefkou:2021fni}, also shown in Fig.~\ref{fig:GFFs}. For the results of this work,the GFFs are shown as obtained using an EMT operator with 2 steps of stout smearing, but consistent results are obtained with 3 steps. 
}
\label{fig:pions_only}
\end{figure*}

The final results for $A_g(t)/A_g(0)$ and $D_g(t)/A_g(0)$ of the quenched pion are shown in Fig.~\ref{fig:pions_only}, and compared against the pion GFFs of Ref.~\cite{Pefkou:2021fni}. The two sets of results are in good agreement.

\end{document}